\documentclass[a4paper,11pt]{article}
\pdfoutput=1 

\usepackage{jinstpub} 
\usepackage{booktabs} 
\usepackage{tabularx}
\usepackage{array}
\usepackage{placeins}
\newcolumntype{L}[1]{>{\raggedright\arraybackslash}p{#1}}
\newcolumntype{Y}{>{\raggedright\arraybackslash}X}
\newcommand{\AFedit}[1]{\textcolor{black}{#1}}

\title{\boldmath Future Muon Physics Experiments: Muon Beams and Experimental Apparatus}

\author[a,b,1]{Yi~Yuan,}
\author[c,1]{Leyun~Gao,\note{Co-first author.}}
\author[a,b,2]{Jian~Tang,\note{Corresponding authors.}}
\author[c,2]{and Qiang~Li}

\affiliation[a]{School of Physics, Sun Yat-sen University, \\ Guangzhou 510275, China}
\affiliation[b]{Platform for Muon Science and Technology, Sun Yat-sen University, \\ Guangzhou 510275, China}
\affiliation[c]{School of Physics and State Key Laboratory of Nuclear Physics and Technology,
                Peking University, \\ Beijing 100871, China}

\emailAdd{tangjian5@mail.sysu.edu.cn}
\emailAdd{qliphy0@pku.edu.cn}

\abstract{Muons are powerful probes of fundamental physics at the precision and intensity frontiers. We review the status and planned upgrades of selected muon-beam facilities worldwide, together with the experimental apparatus \AFedit{associated with representative experiments} across three categories. Charged-lepton flavor violation (CLFV) searches test charged-lepton flavor conservation through rare muon decays, muon-to-electron conversion in nuclei, and muonium--antimuonium conversion. Precision measurements of the muon magnetic and electric dipole moments, muonium spectroscopy, and muonic-atom spectroscopy test the Standard Model and bound-state QED and determine fundamental constants and nuclear charge and magnetization distributions. \AFedit{Muon-scattering programs span low-momentum elastic lepton--proton scattering with PSI MUSE to study the proton radius, lepton universality, and two-photon exchange; high-energy MUonE measurements of hadronic vacuum polarization; NA64$\mu$ missing-momentum searches for invisible dark sectors;} and the phased PKMu program, which investigates muonphilic dark matter, light mediators, LFV muon--electron processes, and atomic effects in resonant annihilation. We emphasize the beam intensity, timing, purity, and phase-space control, as well as beamline and detector design, background suppression, and experimental techniques needed to achieve next-generation sensitivity and precision.}


\arxivnumber{} 

\proceeding{Special Issue: ICFA Beam Dynamics Newsletter\\
            Future Muon Physics Experiments} 

\begin{document}
\maketitle
\flushbottom

\section{Introduction}
\label{sec:intro}

Muons play an essential role at the precision and intensity frontiers of particle and nuclear physics. Their mass, approximately 207 times that of the electron, enhances sensitivity to mass-dependent effects in precision measurements and searches for physics beyond the Standard Model~\cite{Keshavarzi:MuonG2Review2021,Hertzog:Hoferichter2026}. At the same time, their $2.2~\mathrm{\upmu s}$ proper lifetime~\cite{PDG2024} permits the production, transport, and stopping of intense muon beams in dedicated experiments. For relativistic muons, Lorentz time dilation further increases the laboratory-frame decay length, enabling extended accelerator beamlines.

These features support a broad and rapidly evolving program, organized here according to its principal physics question or experimental method. The first category comprises charged-lepton flavor violation (CLFV) searches, which test charged-lepton flavor conservation through rare muon decays, muon-to-electron conversion in nuclei, and muonium--antimuonium conversion. Current searches set upper limits on flavor-changing processes; representative experiments are MACE~\cite{Bai:SnowmassMACE2022,Bai:MACE2024}, Mu2e~\cite{Bernstein:Mu2e2019}, COMET~\cite{Kuno:COMET2013}, Mu3e~\cite{Mu3e:TDR2020}, and MEG~II~\cite{MEGII:Result2025}. The second category consists of flavor-conserving precision measurements of free and bound muons, including the anomalous magnetic moment ($g-2$), a possible muon electric dipole moment (EDM), and the spectra of muonium and muonic atoms. These observables test the Standard Model and bound-state QED and constrain fundamental constants and nuclear structure. The third category comprises muon-scattering experiments \AFedit{over different momentum scales, including PSI MUSE, which compares low-momentum elastic $e^{\pm}p$ and $\mu^{\pm}p$ scattering to study the proton radius, lepton universality, and two-photon exchange; MUonE's high-energy elastic $\mu e$ measurement of hadronic vacuum polarization;} NA64$\mu$ missing-momentum searches for invisibly decaying dark-sector particles\AFedit{;} and PKMu searches for muonphilic dark matter, light mediators, LFV interactions, and bound-electron effects in resonant muon--electron annihilation~\AFedit{\cite{Gilman:MUSE2017,Abbiendi:MUonELOI2019,Andreev:NA64muPRL2024,Andreev:NA64muPRD2024,Liu:PKMuNewPhysics2026,Shen:LFVScalar2026}}.

Although their observables differ, all three categories depend on suitably matched muon beams and detection systems. Their sensitivity is determined by beam intensity, time structure, phase space, polarization, particle purity, and background conditions, together with target, transport, and detector design. Rare positive-muon decay searches require continuous, high-intensity surface-muon beams, such as those available at the Paul Scherrer Institute (PSI), to accumulate statistics~\cite{Prokscha:MuE4PSI2008}. In contrast, conversion experiments use pulsed, low-momentum $\mu^-$ beams stopped in dedicated targets, with delayed observation windows and inter-pulse extinction suppressing prompt beam-related backgrounds~\cite{Bartoszek:Mu2eTDR2015,COMET:PhaseITDR2020}. Spin-precession measurements require storage rings or reaccelerated beams to control systematic effects~\cite{Stratakis:MuonCampus2017,Shimomura:JPARCMUSE2024}. \AFedit{Scattering experiments use PSI's $\pi$M1 beamline for low-momentum lepton--proton scattering with MUSE, the high-energy CERN SPS M2 beam for MUonE and NA64$\mu$, and prospective GeV-scale HIAF beams for PKMu~\cite{Gilman:MUSE2017,Abbiendi:MUonELOI2019,Andreev:NA64muPRL2024,Xu:HIAFMuon2025}. Their detector systems are tailored to each observable, including tracking and time-of-flight particle identification for MUSE, angular tracking for MUonE, and momentum reconstruction plus vetoes for missing-momentum searches~\cite{Gilman:MUSE2017,Abbiendi:MUonELOI2019,Andreev:NA64muPRL2024,Wang:LightDarkSectors2026}}. Progress across the program therefore requires coordinated advances in beam delivery and experimental apparatus.

With several experiments approaching operation and next-generation programs entering construction or design, a review of their beam and detector requirements is timely. This article reviews \AFedit{selected ongoing and planned} muon physics experiments\AFedit{,} with emphasis on beam facilities and experimental systems\AFedit{. The experiments discussed here are representative examples rather than an exhaustive survey of all muon physics programs}. Section~\ref{sec:beams} describes the status and planned upgrades of muon-beam facilities. Sections~\ref{sec:CLFV}--\ref{sec:scattering} cover CLFV searches, precision measurements with free and bound muons, and high-precision muon-scattering experiments, respectively. Section~\ref{sec:summary} provides a summary and outlook.

\section{Status and upgrades of muon beam facilities}
\label{sec:beams}

Most existing and planned muon beam facilities are driven by high-intensity proton accelerators. In such facilities, a proton beam impinges on a production target, producing pions that subsequently decay into muons. Depending on the production mechanism and subsequent processes, the muon beams are classified as surface, decay (cloud), or slow/ultra-slow muons, as well as high-energy muon beams ranging from a few GeV to several hundred GeV. According to the time structure of the primary proton beam, they are further categorized as continuous-wave (CW) or pulsed muon beams.


Several muon beam facilities are currently in operation or under construction worldwide. Among them, the Swiss \AFedit{Research Infrastructure for Particle Physics (CHRISP)} at the Paul Scherrer Institute (PSI) and the Centre for Molecular and Materials Science (CMMS) at TRIUMF are representative \AFedit{facilities operating with continuous muon beams, providing surface and decay muon beams for different research purposes. CHRISP mainly supports muonium research, rare-process searches, and precision measurements, while CMMS serves as a platform for muon spin rotation/relaxation/resonance ($\mu$SR) studies and muon spectroscopy~\cite{PSI:CHRISP,TRIUMF:MuonBeamlines}}. The ISIS Neutron and Muon Source at the Rutherford Appleton Laboratory (RAL) and the J-PARC Muon Science Establishment (J-PARC MUSE) represent pulsed muon beam facilities, whose time structure is advantageous for muon spectroscopy, muonic atom studies, and time-resolved precision measurements~\cite{Hillier:ISIS2019,Miyake:JPARCMUSE2012}.

In particular, the J-PARC MUSE H-line is dedicated to high-statistics measurements of the muonium and muonic helium hyperfine splitting (HFS)~\cite{Strasser:MuSEUM2025}, while the associated ultra-slow muon beamline and reacceleration technologies at J-PARC also serve the muon $g-2$ and EDM experimental programs.

The Fermilab Muon Campus was developed to support the Muon $g-2$ and Mu2e programs~\cite{Stratakis:MuonCampus2017}. It previously delivered polarized $3.094~\mathrm{GeV}/c$ $\mu^+$ beams to Muon $g-2$. The facility has since been reconfigured to deliver slow-extracted $8~\mathrm{GeV}$ proton spills for Mu2e, with commissioning runs conducted in 2025 and early 2026~\cite{Nagaslaev:Mu2eSlowExtraction2026}. The CERN SPS M2 beamline provides high-energy muon beams with momenta of order $160~\mathrm{GeV}/c$, serving as an essential beamline infrastructure for fixed-target scattering and missing-momentum experiments including COMPASS/AMBER~\cite{Adams:AMBERProposal2019}, MUonE~\cite{Abbiendi:MUonE2017}, and NA64$\mu$~\cite{Andreev:NA64muPRD2024}. In addition, the Muon Science Innovative Channel (MuSIC) at the Research Center for Nuclear Physics (RCNP), Osaka University~\cite{Cook:MuSIC2017}, employs a pion-capture-solenoid scheme for high-acceptance muon production and transport. It was the first time to demonstrate the feasibility of muon capture inside the solenoid and offer an important reference for future high-intensity muon beamline design.
This review covers only the operational modes directly relevant to future muon physics experiments. For comprehensive information about these facilities, please refer to Ref.~\cite{Chen:muoniumreview-2026}.

To meet the demands for higher intensity and more stringent beam control, several institutions are pursuing upgrades to existing facilities and constructing dedicated beamline systems for specific experiments. At PSI, the High-Intensity Muon Beam (HIMB) project is designed to deliver a surface muon intensity of $\mathcal{O}(10^{10})~\mu^+/\mathrm{s}$, and serve experiments including Mu3e Phase~II, muonium spectroscopy, $\mu$SR, and other high-statistics low-energy muon experiments~\cite{DalMaso:HIMB2023}. For $\mu^-N\to e^-N$ conversion searches, the Fermilab Muon Campus is being commissioned for Mu2e~\cite{Bartoszek:Mu2eTDR2015,Nagaslaev:Mu2eSlowExtraction2026}, while J-PARC is constructing a dedicated beamline system for COMET~\cite{COMET:PhaseITDR2020}. The key performance parameters include the pulsed proton beam structure, momentum and charge selection, beam extinction, and beam-related background conditions. Mu2e employs a superconducting solenoid system composed of production, transport, and detector solenoids arranged in sequence. COMET adopts a phased curved-solenoid transport scheme, with Phase~I designed for an $8~\mathrm{GeV}$, $3.2~\mathrm{kW}$ proton beam, and Phase~II assuming a higher beam power of $56~\mathrm{kW}$. The Fermilab PIP-II project, featuring an $800~\mathrm{MeV}$ superconducting radio-frequency linac, will enhance the overall beam delivery capability of the accelerator complex and provide the foundation for Mu2e-II upgrade studies in the PIP-II era~\cite{Pathak:PIPII2024}.

\begin{table}[t]
\centering
\caption{Selected muon-beam facilities, their status, and representative performance. Intensity values retain the definitions and reference locations used in the cited sources and are therefore not directly comparable.}
\label{tab:facilities}

\footnotesize
\setlength{\tabcolsep}{3.5pt}
\renewcommand{\arraystretch}{1.10}

\begin{tabularx}{\linewidth}{@{}%
                             >{\raggedright\arraybackslash}p{2.55cm}%
                             >{\raggedright\arraybackslash}p{2.05cm}%
                             >{\raggedright\arraybackslash}p{4.60cm}%
                             >{\raggedright\arraybackslash}X@{}}
\toprule
Facility / beamline
& Status (July 2026)
& Representative beam (momentum / energy)
& Representative intensity scale \\
\midrule

\AFedit{PSI $\pi$E5 (CHRISP)%
~\cite{PSI:PiE5,MEG:Polarization2016,MEGII:Design2018}}
& \AFedit{Operating}
& \AFedit{Surface $\mu^+$ ($\sim28~\mathrm{MeV}/c$);
cloud $\mu^\pm$ (beamline range $10$--$120~\mathrm{MeV}/c$)}
& \AFedit{$\sim10^8~\mu^+/\mathrm{s}$ for surface muons} \\

TRIUMF CMMS%
~\cite{Kreitzman:TRIUMF2018,Kreitzman:TRIUMF2025}
& Operating
& Surface $\mu^+$ ($\sim29~\mathrm{MeV}/c$);
decay $\mu^\pm$ ($40$--$120~\mathrm{MeV}/c$)
& $\sim10^6~\mu^+/\mathrm{s}$ for surface muons \\

ISIS at RAL%
~\cite{Hillier:ISIS2019}
& Operating
& Pulsed surface $\mu^+$ ($28~\mathrm{MeV}/c$);
decay $\mu^\pm$ ($17$--$90~\mathrm{MeV}/c$)
& $\sim10^5$--$10^6~\mu/\mathrm{s}$ \\

J-PARC MUSE%
~\cite{Miyake:JPARCMUSE2012,Shimomura:JPARCMUSE2024}
& Operating
& Pulsed surface $\mu^+$ ($\sim30~\mathrm{MeV}/c$);
decay $\mu^\pm$ ($5$--$120~\mathrm{MeV}/c$);
ultra-slow $\mu^+$ ($50~\mathrm{eV}$--$30~\mathrm{keV}$)
& $\sim10^7~\mu/\mathrm{s}$ conventional;
$\sim10^2~\mu^+/\mathrm{s}$ ultra-slow \\

RCNP MuSIC%
~\cite{Cook:MuSIC2017,Cook:MuSICProduction2013}
& Operating
& High-acceptance low-energy $\mu^\pm$
($\sim28$--$110~\mathrm{MeV}/c$)
& $(4.2\pm1.1)\times10^8~\mu/\mathrm{s}$
($\mu^++\mu^-$; 400-W equivalent; extrapolated) \\

CERN SPS M2%
~\cite{Abbon:COMPASS2007,Abbiendi:MUonELOI2019}
& Operating
& High-energy $\mu^\pm$
($150$--$160~\mathrm{GeV}/c$)
& $\sim1.3\times10^7~\mu/\mathrm{s}$
(MUonE-type) \\

\midrule

PSI HIMB%
~\cite{DalMaso:HIMB2023,Valetov:HIMBBeamline2024}
& Upgrade
& High-intensity surface $\mu^+$
(near $29~\mathrm{MeV}/c$)
& $\mathcal{O}(10^{10})~\mu^+/\mathrm{s}$
(design) \\

Fermilab Mu2e%
~\cite{Mu2e:RunISensitivity2023,Nagaslaev:Mu2eSlowExtraction2026}
& Commissioning
& Pulsed low-momentum $\mu^-$
($\sim50~\mathrm{MeV}/c$); Al stopping target
& Muon stopping rate in Al:
$\sim10^{10}\,\mathrm{s}^{-1}$
(design simulation) \\

J-PARC COMET Phase~I%
~\cite{COMET:PhaseITDR2020}
& Construction
& Pulsed low-momentum $\mu^-$
($\sim40~\mathrm{MeV}/c$); Al stopping target
& Muon stopping rate in Al:
$1.2\times10^9\,\mathrm{s}^{-1}$
(design simulation) \\

CSNS MELODY%
~\cite{Chen:MELODYBeamline2023,Zhang:MELODYIntensity2025}
& Construction
& Pulsed surface $\mu^+$
(near $29~\mathrm{MeV}/c$)
& $\sim10^5~\mu/\mathrm{pulse}$ at $1~\mathrm{Hz}$;
goal $\sim10^6~\mu/\mathrm{pulse}$ \\

HIAF--HFRS%
~\cite{Xu:HIAFMuon2025,Sheng:HFRSOptics2024}
& Trial operation;
muon-mode feasibility
& Secondary $\mu^\pm$
($1.5$--$3.5~\mathrm{GeV}/c$)
& $\mathcal{O}(10^6)~\mu/\mathrm{s}$ unpurified;
$\mathcal{O}(10^5)~\mu/\mathrm{s}$
rigidity-purified (simulated) \\

CiADS muon source%
~\cite{Cai:CiADS2024}
& Design / R\&D
& High-intensity surface $\mu^+$
(near $29~\mathrm{MeV}/c$)
& Phase~I ($\sim300~\mathrm{kW}$):
$\sim10^8~\mu^+/\mathrm{s}$ (design) \\

SHINE muon source%
~\cite{Liu:SHINE2025,Lv:SHINEMuonSource2023}
& Feasibility study
& Electron-driven surface $\mu^+$
(near $29~\mathrm{MeV}/c$)
& $\sim3\times10^6~\mu^+/\mathrm{s}$
transported (simulated, $50~\mathrm{kHz}$) \\

\bottomrule
\end{tabularx}
\end{table}
\FloatBarrier

Beyond the aforementioned upgrades, a number of facilities under construction, in planning, or at the conceptual study stage are broadening the range of applicability of muon beams. In China, CSNS is constructing MELODY (Muon station for sciEnce technoLOgy and inDustrY), the first surface muon source in China, operating in a pulsed mode at $1~\mathrm{Hz}$, which will complement the domestic low-energy muon experimental infrastructure~\cite{Chen:MELODYBeamline2023,Bao:MELODYProgress2023}. In parallel, two muon beamline initiatives have been proposed~\cite{An:HuizhouPrecision2025}. The HIAF facility entered trial operation in July 2026, while the use of HFRS for GeV-scale secondary muon beams remains a simulation-based feasibility study~\cite{Xu:HIAFMuon2025}. In August 2026, the HIAF muon-source team used a $1.4~\mathrm{GeV}/c$ mixed proton--muon beam at HIAF to demonstrate non-destructive imaging of a $30$-mm Ti--SS--Ti sample in a roughly $400$-min muon-tomography run~\cite{He:NuFact2026}. Its prospective high-purity beam and controlled phase space would support compact PKMu fixed-target and scattering detectors, for which tracking acceptance, target material, and veto efficiency are central design parameters~\cite{Zhang:MuonDMBeam2026,Wang:LightDarkSectors2026,Shen:LFVScalar2026}. In parallel, the CiADS project plans a Phase~I surface muon beam driven by an approximately $300~\mathrm{kW}$ proton beam, with a design intensity of order $10^8~\mu^+/\mathrm{s}$ for $\mu$SR and low-energy muon experiments~\cite{Cai:CiADS2024}. Furthermore, the SHINE facility is investigating a surface muon scheme driven by a high-repetition-rate electron beam. Simulations with an $8~\mathrm{GeV}$, $100~\mathrm{pC}$ electron bunch on a copper target indicate a transported surface muon rate of about $3\times10^6~\mu^+/\mathrm{s}$ at $50~\mathrm{kHz}$~\cite{Lv:SHINEMuonSource2023}.

Around the world, similar concepts driven by high-energy electron beams are also being considered. A Jefferson Lab white paper on beam-dump facilities discusses secondary muon beams produced when the electron beam from the Continuous Electron Beam Accelerator Facility (CEBAF) impinges on a beam dump, with a simulated intensity of $\mathcal{O}(10^8)~\mu/\mathrm{s}$ and an energy spectrum extending to several GeV~\cite{Achenbach:JLabBeamDump2025}. Studies at ORNL/SNS on the proposed Single Event Effects and Muon Spectroscopy (SEEMS) facility and on laser-assisted proton pulse extraction explore the possibility of a co-located muon spectroscopy source with narrow time structure~\cite{Williams:SEEMS2022}. In Korea, related studies at RAON have proposed an antimuon beamline design based on the heavy-ion facility~\cite{Choi:RAONMuon2014}. In addition, slow muon, ultra-slow muon and reaccelerated muon techniques will improve emittance and phase-space control for critical beam conditions in the near future.

These facilities and beamlines are complementary in intensity, time structure, momentum range, emittance, and particle purity, thereby providing different beam conditions for the three classes of experiments discussed below. The development of muon beam facilities and experimental apparatus will collectively drive further improvements in the physics prospects. Table~\ref{tab:facilities} summarizes representative beam modes, momentum or energy coverage, and characteristic intensity scales, spanning low-energy surface muons to GeV- and hundred-GeV-scale high-energy muon beams.

\FloatBarrier

\section{Charged-lepton flavor violation searches with muons}
\label{sec:CLFV}

\begin{figure}[htbp]
\centering
\includegraphics[width=1\textwidth]{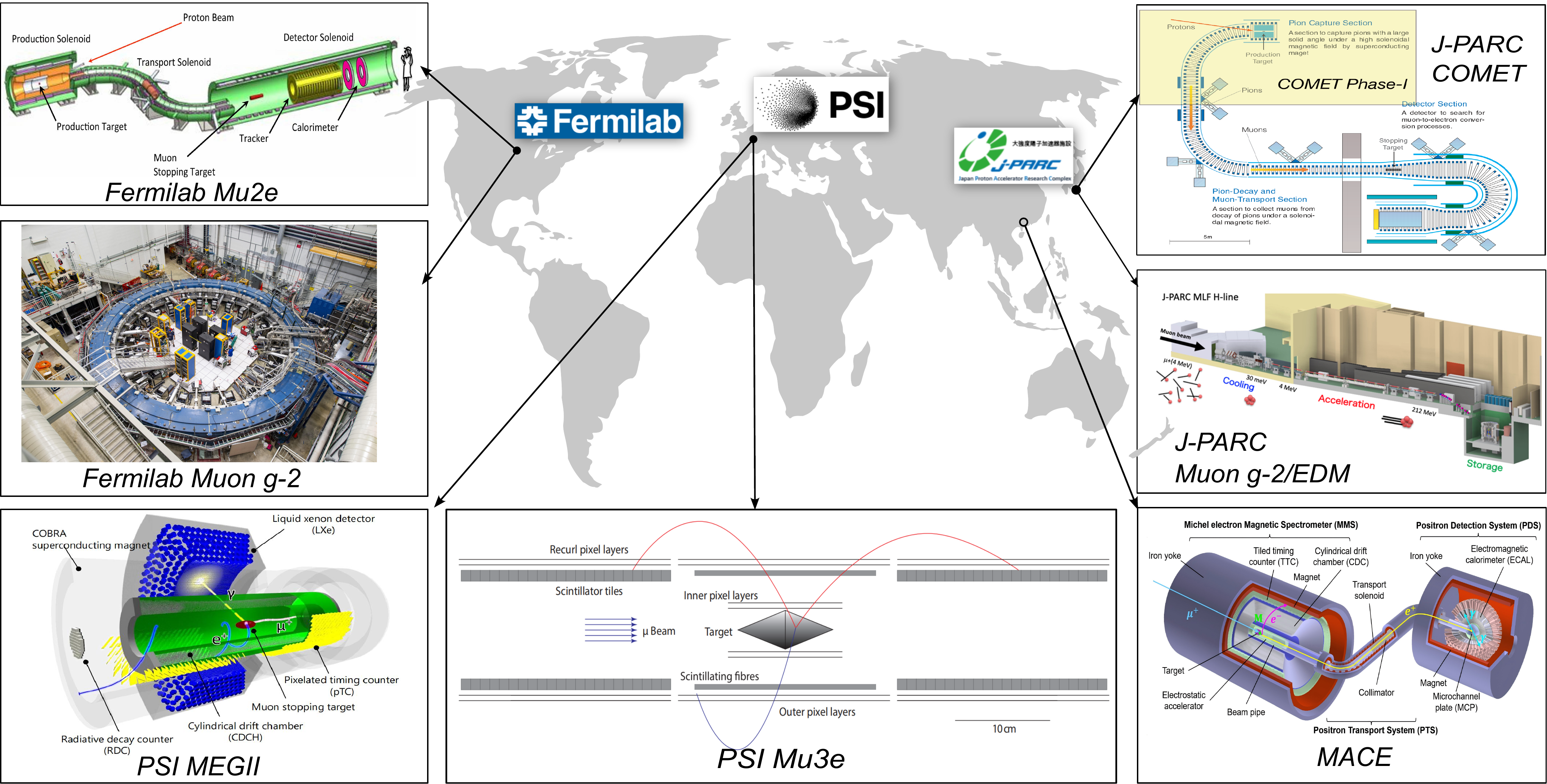}
\caption{Geographic distribution of selected representative current and planned muon experiments discussed in this review. Only the main CLFV and free-muon precision experiments are shown. The apparatus schematics and photographs are taken from the design reports, detector papers, or official experiment sources of Mu2e~\cite{Bartoszek:Mu2eTDR2015}, Muon $g-2$ (photograph by Reidar Hahn, Fermilab)~\cite{Fermilab:MuonG2Photo2021}, MEG~II~\cite{MEGII:Design2018}, Mu3e~\cite{Mu3e:TDR2020}, COMET~\cite{COMET:PhaseITDR2020}, J-PARC $g-2$/EDM~\cite{Abe:JPARCG2EDM2019}, and MACE~\cite{Bai:MACE2024}.}
\label{fig:facilities-map}
\end{figure}

It is one of the most interesting scenarios to search for charged-lepton flavor violation with muons, including muonium--antimuonium conversion, $\mu^+\to e^+\gamma$, $\mu^+\to e^+e^-e^+$, and $\mu^-N\to e^-N$~\cite{Palo:CLFV2025}. Figure~\ref{fig:facilities-map} provides a geographic overview of selected current and planned muon experiments discussed in this review, spanning the CLFV searches of this section and the free-muon precision measurements of Section~\ref{sec:precision}.

The signals and backgrounds differ among the four channels, leading to different beam and detector designs. Their sensitivities depend mainly on muon statistics, detector performance, and background control.

\subsection{Searches for muonium--antimuonium conversion}
\label{sec:mace}

The spontaneous conversion of muonium ($\mu^+e^-$) to antimuonium ($\mu^-e^+$) is a bound-state CLFV search channel whose signal event is tagged by the coincidence of a Michel electron and a low-energy atomic positron from antimuonium decay~\cite{Han:DoublyChargedHiggs2021}. MACE (Muonium-to-Antimuonium Conversion Experiment) plans to inject a surface muon beam of approximately $10^8~\mu^+/\mathrm{s}$ into a porous, perforated silica aerogel target to produce muonium, which then diffuses into vacuum. If conversion occurs, the antimuonium decay event can be identified by this coincidence signature~\cite{Bai:MACE2024,Zhao:Aerogel2024}.

The MACE detector concept consists of three subsystems: Michel electron Magnetic Spectrometer (MMS), Positron Transport System (PTS), and Positron Detector System (PDS). The MMS reconstructs the Michel electron trajectory using a cylindrical drift chamber (CDC) and provides $\mathcal{O}(100)~\mathrm{ps}$ decay-time tagging via a tiled timing counter (TTC). The PTS accelerates and transports the low-energy atomic positron through an electrostatic accelerator and a solenoid beamline. The PDS comprises the MCP and the near-$4\pi$ CsI(Tl) electromagnetic calorimeter (ECAL): the MCP records the positron arrival position and time, and the ECAL detects the back-to-back $511~\mathrm{keV}$ $\gamma$-rays from positron annihilation. The final signal selection requires a triple coincidence among the electron track in the MMS, the positron hit on the MCP, and the annihilation $\gamma$-rays in the ECAL~\cite{Lu:MACEPTS2025,Chen:MACECalorimeter2024}.

MACE aims at a sensitivity of $\mathcal{O}(10^{-13})$ at 90{\%} confidence level, more than two orders of magnitude improvement beyond the previous experiment at PSI~\cite{Willmann:MACS1999,Bai:MACE2024}.
MACE Phase~I will validate the detector technology and enable pilot searches for rare muon and muonium processes, including $\mathrm{M}\to\gamma\gamma$ and $\mu^+\to e^+\gamma\gamma$, with sensitivities at the $\mathcal{O}(10^{-12})$ level~\cite{Zhao:MACEProgress2024}. The Phase~I detector is based on an ECAL, supplemented by an inner tracking system consisting of a scintillating-fiber tracker and a hodoscope. A surface muon beam for Phase~I is expected to be provided by future high-intensity muon beam facilities such as CiADS~\cite{Cai:CiADS2024}.

\subsection{\texorpdfstring{Searches for $\mu^-N\to e^-N$}{Searches for mu-N to e-N}}
\label{sec:mu-e-conversion}

Searching for $\mu^-N\to e^-N$ focuses on the coherent neutrinoless conversion of a $\mu^-$ bound in a muonic atom. The signal is a monoenergetic electron whose energy depends on the target nucleus; for aluminum it is approximately $105~\mathrm{MeV}$. The dominant backgrounds include decay-in-orbit (DIO) electrons, radiative pion capture (RPC), cosmic rays, and inter-pulse beam leakage. The current best limit is $R_{\mu e}<7\times10^{-13}$ at 90\% CL, set by SINDRUM~II on a gold target~\cite{Bertl:SINDRUMII2006}. Mu2e and COMET use aluminum stopping targets and aim for a sensitivity of $\mathcal{O}(10^{-17})$~\cite{Bernstein:Mu2e2019,COMET:PhaseITDR2020}. Both experiments use pulsed low-momentum $\mu^-$ beams and solenoidal transport systems that integrate pion/muon production, momentum and charge selection, the stopping target, and conversion-electron detection into a single apparatus.

The Mu2e experiment at Fermilab employs an integrated superconducting solenoid system consisting of production, transport, and detector solenoids arranged in series. An $8~\mathrm{GeV}$ pulsed proton beam strikes the production target to produce pions and muons. The pulsed time structure and delayed observation window suppress prompt backgrounds, including radiative pion capture and other beam-related backgrounds. Negative muons pass through an S-shaped transport solenoid, where collimators and absorbers select their momentum and charge and reject beam backgrounds, before stopping in aluminum foil targets. The detector sits inside the detector solenoid. At the center is a low-mass straw-tube tracker that measures the $\sim105~\mathrm{MeV}$ electron momentum to separate DIO backgrounds. Downstream is an undoped CsI dual-disk calorimeter for particle identification, track seeding, and timing. A multi-layer veto based on plastic scintillators encloses the detector to reject cosmic-ray-induced backgrounds, while the Stopping Target Monitor measures capture-photon rates to infer the rate and integrated number of $\mu^-$ stops in the Al target. Mu2e aims for a 90\% CL upper limit of $R_{\mu e}<8\times10^{-17}$~\cite{Bartoszek:Mu2eTDR2015}. The experiment is currently in detector installation and commissioning, with Run~I expected to begin in the second half of 2027~\cite{Ricci:Mu2e2025,Nagaslaev:Mu2eSlowExtraction2026}.

The COMET experiment at J-PARC follows a staged program based on a large-acceptance curved-solenoid transport system. Phase~I uses an $8~\mathrm{GeV}$, $3.2~\mathrm{kW}$ proton beam and the first section of the curved-solenoid transport system. It aims for a single-event sensitivity of $3.1\times10^{-15}$ with an expected background of 0.032 events. Its detector comprises two complementary systems. CyDet contains the stopping target, a Cylindrical Drift Chamber (CDC), and a Cylindrical Trigger Hodoscope (CTH), and searches for conversion electrons in the solenoidal magnetic field. StrECAL, composed of a straw-tube tracker and an electromagnetic calorimeter, measures beam-related backgrounds and provides design input for Phase~II. Phase~$\alpha$ has completed proton beamline construction and verified muon transport through the curved solenoid~\cite{Fujii:COMET2023,Oishi:COMETRangeCounter2025,Xu:COMETMBM2024}. Phase~II is designed for a beam power of $56~\mathrm{kW}$. It employs a C-shaped $180^\circ$ curved muon-transport solenoid, followed by a curved electron spectrometer and a detector solenoid that houses a vacuum straw-tube tracker and an electromagnetic calorimeter. The longer muon-transport path and the electron-spectrometer momentum selection further suppress beam-related backgrounds. Phase~II aims for a single-event sensitivity of $2.6\times10^{-17}$~\cite{Moritsu:COMET2022}.

At Fermilab and J-PARC, studies of $\mu^-N\to e^-N$ conversion experiments beyond Mu2e and COMET Phase~II are under way. This channel has lower accidental-coincidence backgrounds than positive muon rare-decay channels and is therefore more scalable with high-intensity negative muon beams. Future experiments may require further optimization of the beam time structure, phase space, and extinction. PRISM/AMF-type phase rotation and phase-space purification could reduce beam-related backgrounds and improve stopping-target conditions~\cite{Palo:CLFV2025}. A PRISM/AMF-type system could deliver lower-momentum negative muon beams with a narrower phase space, allowing a lower-mass stopping target and more precise measurement of the conversion-electron momentum.

\subsection{\texorpdfstring{Searches for $\mu^+\to e^+e^-e^+$}{Searches for mu to eee}}
\label{sec:mu3e}

Mu3e searches for $\mu^+\to e^+e^-e^+$. The signal is an $e^+e^-e^+$ triplet from a $\mu^+$ decay at rest, identified through a common vertex and coincidence in time, with total energy and momentum consistent with a muon decay at rest. The current experimental upper limit is $\mathcal{B}(\mu^+\to e^+e^-e^+)<1.0\times10^{-12}$ at 90\% CL, set by the SINDRUM experiment~\cite{Bellgardt:SINDRUM1988}. One background is the Standard-Model internal-conversion decay $\mu^+\to e^+e^-e^+\nu_e\bar\nu_\mu$, which has the same charged final state but includes missing energy and momentum carried by neutrinos. Energy and momentum measurements distinguish it from the signal. Accidental coincidences form the other principal background and are suppressed by vertex and timing measurements. Because the final-state particles have low momenta, multiple scattering limits the momentum and vertex resolutions, requiring an ultra-low-mass detector.

Mu3e is located at PSI and follows a phased program~\cite{Amarinei:Mu3e2025}. Phase~I uses the existing PSI continuous muon beam at a muon stopping rate of $10^8\,\mathrm{s}^{-1}$ and aims for a single-event sensitivity of $2\times10^{-15}$. Muons stop in a hollow double-cone target at the detector center, and their decay products traverse the pixel and timing layers multiple times in a solenoidal magnetic field. The tracking system uses ultra-thin MuPix HV-MAPS silicon pixel sensors (MuPix11: $50$--$70~\mathrm{\upmu m}$ thick, spatial resolution $\sim23~\mathrm{\upmu m}$) to reconstruct low-momentum tracks with minimal material. Scintillating fibers in the central region provide a timing resolution of approximately $250~\mathrm{ps}$, while scintillating tiles at the recurl stations achieve approximately $40~\mathrm{ps}$. Together with pixel tracking, these timing detectors suppress accidental backgrounds and provide the vertex, timing, and momentum information needed to reconstruct three-track candidates.

Mu3e Phase~I is in detector installation and commissioning and the vertex detector has been installed for the 2026 beamtime. Its continuous muon beam and ultra-low-mass multi-track detector are designed together to control multiple scattering and accidental backgrounds at high stopping rates. Increasing the beam intensity improves the statistical reach but places stricter demands on timing resolution, material budget, data-acquisition bandwidth, and online reconstruction. Phase~II will use a beam provided by HIMB to reach a stopping rate of about $2\times10^9\,\mathrm{s}^{-1}$. Building on the Phase~I layout, which already includes recurl pixel stations and timing detectors, Phase~II will enlarge the detector acceptance and adapt the apparatus to the higher stopping rate, further improving momentum resolution and accidental-background rejection. It aims for a sensitivity of $\mathcal{O}(10^{-16})$~\cite{Aiba:HIMB2021}.

\subsection{\texorpdfstring{Searches for $\mu^+\to e^+\gamma$}{Searches for mu to e gamma}}
\label{sec:meg-ii}

The $\mu^+\to e^+\gamma$ search relies on positron--photon coincidence measurements. For a $\mu^+$ decaying at rest, the signal is a $52.8~\mathrm{MeV}$ positron--photon pair emitted back-to-back from the same stopping-target location and coincident in time. The principal backgrounds are accidental coincidences of Michel decay positrons with uncorrelated photons, and genuine radiative muon decays in which the neutrinos carry very little momentum. Accordingly, the sensitivity depends critically on the resolution of positron momentum, photon energy, relative angle, and relative timing.

MEG~II operates at PSI with a quasi-continuous positive muon beam and a design stopping rate of approximately $7\times10^7\,\mathrm{s}^{-1}$ in a thin target; during the 2021--2022 data taking the realized stopping rate was typically about $3\times10^7\,\mathrm{s}^{-1}$ (within $2$--$5\times10^7\,\mathrm{s}^{-1}$). A low-mass single-volume cylindrical drift chamber (CDCH) reconstructs the positron in a gradient magnetic field. A pixelated timing counter (pTC) measures the positron time. A liquid-xenon (LXe) calorimeter measures the photon energy, time, and entrance position. Offline Kalman-filter reconstruction extrapolates the positron trajectory to the stopping target. The photon direction is obtained from the reconstructed interaction point in the LXe calorimeter and the decay vertex on the target. These measurements define the positron--photon coincidence. Compared with MEG, MEG~II uses a high-granularity CDCH to reduce multiple scattering and replaces the inner-face PMTs in the LXe calorimeter with MPPCs to improve photon position reconstruction. The resulting design sensitivity is $6\times10^{-14}$~\cite{MEGII:Design2018}. Based on 2021--2022 data~\cite{MEGII:Result2025}, the experiment achieved a sensitivity of $2.2\times10^{-13}$ and set an upper limit of $\mathcal{B}(\mu^+\to e^+\gamma)<1.5\times10^{-13}$ at 90\% CL.

The central challenge beyond MEG~II is the approximate quadratic scaling of accidental background with the muon stopping rate. The existing calorimetric approach and gas drift chamber would face excessive occupancy and backgrounds at muon stopping rates $R_\mu\gtrsim10^9\,\mathrm{s}^{-1}$. Photon conversion could improve photon energy and directional resolution, while a more rate-tolerant silicon tracker may be needed for positron reconstruction. These concepts aim to improve the sensitivity by roughly one order of magnitude relative to MEG~II~\cite{Cattaneo:FutureMEGLoI2025}.

\begin{figure}[htbp]
\centering
\includegraphics[width=1\textwidth]{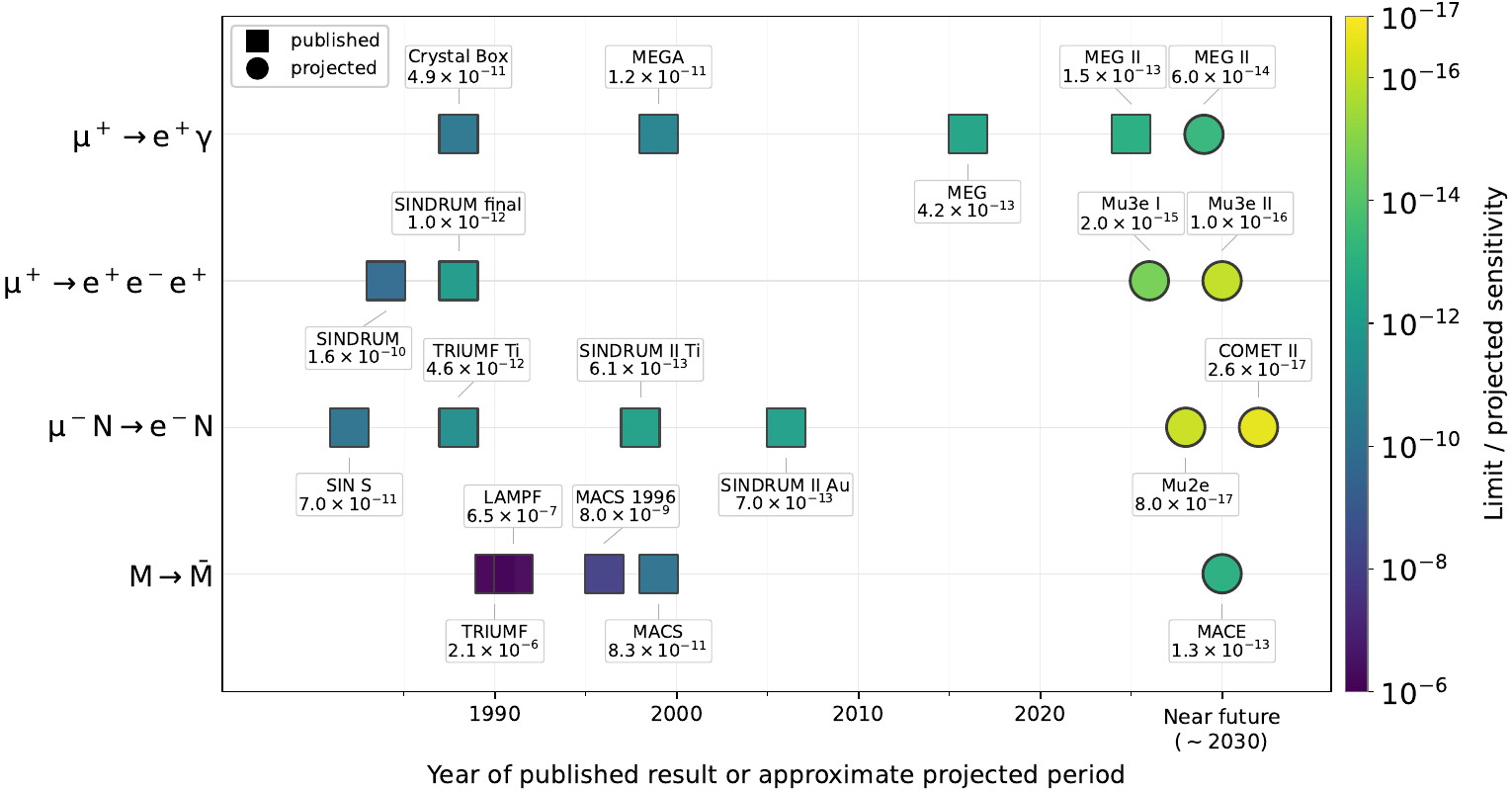}
\caption{Evolution of experimental sensitivities in the four muon CLFV channels discussed in this section.
         The numerical values refer to channel-specific observables:
         the branching ratios for $\mu^+\to e^+\gamma$ and
         $\mu^+\to e^+e^-e^+$, the conversion rate $R_{\mu e}$ for $\mu^-N\to e^-N$, and the conversion probability $P(\mathrm{M}\to\overline{\mathrm{M}})$ for muonium--antimuonium conversion.
         Marker color encodes the magnitude of the quoted limit or sensitivity.
         Square markers denote published upper limits, whereas circular markers denote projected sensitivities as quoted by the respective experiments. The values used in this figure are compiled in Ref.~\cite{Chen:HistoryPlot} and are taken from published or expected limits~\cite{Bellgardt:SINDRUM1988,Willmann:MACS1999,Bertl:SINDRUMII2006,MEGII:Design2018,Mu3e:TDR2020,Aiba:HIMB2021,Bartoszek:Mu2eTDR2015,COMET:PhaseITDR2020,Moritsu:COMET2022,Bai:MACE2024,MEGII:Result2025}.}
\label{fig:clfv-sensitivity}
\end{figure}

Figure~\ref{fig:clfv-sensitivity} summarizes the sensitivity evolution across the four CLFV channels over the recent decades. In the $\mu^+\to e^+\gamma$ channel, the limit has improved from $\mathcal{O}(10^{-11})$ (MEGA) to $1.5\times10^{-13}$ (MEG~II), with the full MEG~II dataset expected to reach $6\times10^{-14}$. For $\mu^+\to e^+e^-e^+$, Mu3e Phase~I and~II aim to improve upon the SINDRUM bound by three to four orders of magnitude. The $\mu^-N\to e^-N$ channel is poised for the largest single-step gain: Mu2e and COMET Phase~II aim for sensitivities at $\mathcal{O}(10^{-17})$, roughly four orders of magnitude beyond SINDRUM~II. In the muonium--antimuonium sector, MACE will extend the reach by more than two orders of magnitude relative to MACS.

The projected gains require improvements in both muon beams and detector performance. In positive muon decay searches, higher stopping rates increase accidental coincidences and detector occupancy, placing stricter demands on timing resolution, material budget, and online event selection. Conversion experiments instead depend strongly on beam extinction, beam-related background control, and conversion-electron momentum resolution. Further sensitivity gains therefore require better beam phase-space control, low-mass high-granularity tracking, high-resolution calorimetry, and higher-bandwidth data-acquisition and trigger systems.

\section{Precision measurements with muons, muonium and muonic atoms}
\label{sec:precision}

This section covers flavor-conserving precision observables: the muon anomalous magnetic moment and electric dipole moment, muonium spectroscopy, and muonic-atom spectroscopy. Muonium--antimuonium conversion is instead treated in Section~\ref{sec:CLFV}, because it is a CLFV search even though it uses the same bound state. Measurements of the anomalous magnetic moment test the Standard Model, and searches for an electric dipole moment probe possible CP violation. Muonium spectroscopy tests bound-state QED and determines fundamental constants, while muonic atom spectroscopy constrains nuclear charge and magnetization distributions.

\subsection{Muon magnetic and electric dipole moments}
\label{sec:g-2}

Measurements of the muon dipole moments rely on the spin precession of polarized muons in controlled electromagnetic fields, read out through the temporal, energy, or spatial distributions of decay positrons. Fermilab Muon $g-2$ has measured the anomalous magnetic moment $a_\mu=(g_\mu-2)/2$ \AFedit{and set a direct upper limit on the muon electric dipole moment (EDM).} J-PARC $g-2$/EDM is designed to measure $a_\mu$ and search for an EDM, \AFedit{while PSI muEDM focuses on the EDM search}.

Fermilab Muon $g-2$ follows the storage-ring approach of BNL E821. Polarized positive muons circulate in a $1.45~\mathrm{T}$ ring, and the time and energy distributions of decay positrons determine the anomalous spin-precession frequency $\omega_a$. The magnetic field is measured through the corresponding proton precession frequency, allowing $a_\mu$ to be obtained from the two frequencies. The main apparatus systems are the storage-ring magnet, injection and collimation systems, electromagnetic calorimeters, straw trackers, and magnetic field monitors~\cite{Grange:MuonG2TDR2015}. The analysis accounts for uncertainties from the extraction of $\omega_a$, beam-dynamics corrections, and the determination of the muon-weighted magnetic field. The experiment collected data over six run periods from 2018 to 2023 and reported its final result for $a_\mu$ in 2025 with a precision of $127~\mathrm{ppb}$. The combined Fermilab--BNL experimental world average reaches $124~\mathrm{ppb}$~\cite{MuonG2:Measurement2025,MuonG2:Final2026}. \AFedit{Using the 2019--2020 data, the collaboration also searched for a muon EDM through oscillations in the mean vertical decay angle of positrons measured by the straw trackers. The result is consistent with zero and sets a direct upper limit of $|d_\mu|<1.10\times10^{-19}\,e\,\mathrm{cm}$ at 95\% confidence level~\cite{MuonG2:EDM2026}}.

A recent study has outlined a possible follow-up concept for Fermilab Muon $g-2$ that would retain the existing storage ring and beamlines, profiting from the higher proton flux and faster repetition cycle expected from the PIP-II linac. Together with accelerator and storage-ring upgrades, the concept aims for a precision near $40~\mathrm{ppb}$ on $a_\mu$ through a roughly ten-fold increase in statistics and reduced systematic uncertainties~\cite{Hertzog:Hoferichter2026}.

J-PARC $g-2$/EDM and PSI muEDM pursue independent measurement routes with different detector configurations and systematic uncertainties. The J-PARC $g-2$/EDM experiment will use a reaccelerated thermal muon beam, whose emittance is about three orders of magnitude below that of a conventional beam, allowing the muons to be stored at much lower momentum in a compact magnet rather than a large storage ring. The spin information will be extracted by reconstructing tracks of individual decay positrons. The baseline design targets a statistical uncertainty of $450~\mathrm{ppb}$ on $a_\mu$, with systematic uncertainties below $70~\mathrm{ppb}$, and an EDM statistical sensitivity of $1.5\times10^{-21}\,e\,\mathrm{cm}$~\cite{Abe:JPARCG2EDM2019,JPARC:g2EDMStatus2025}. \AFedit{Higher-sensitivity upgrades are under study, including an increase in the beam momentum from $300$ to $600~\mathrm{MeV}/c$ and in the storage magnetic field from $3$ to $6~\mathrm{T}$. Assuming $5.7\times10^{11}$ detected positrons and a beam polarization of 50\%, this configuration could reduce the statistical uncertainty on $a_\mu$ to below $120~\mathrm{ppb}$. Increasing the polarization to 70\% could further reduce this uncertainty to below $90~\mathrm{ppb}$~\cite{JPARC:g2EDMStatus2025}.}

PSI muEDM is designed to use the frozen-spin technique, in which the electromagnetic field configuration suppresses the dominant magnetic moment precession so that a small EDM-induced spin rotation can be read out through the decay positron asymmetry. The Phase~I compact frozen-spin trap will use muons with momenta of about $23~\mathrm{MeV}/c$ in a $2.5~\mathrm{T}$ solenoid, corresponding to an effective electric field of about $165~\mathrm{MV}/\mathrm{m}$ in the muon rest frame, and aims for a sensitivity of $4\times10^{-21}\,e\,\mathrm{cm}$. A later Phase~II with a dedicated apparatus and a higher-momentum beamline targets $6\times10^{-23}\,e\,\mathrm{cm}$. The compact storage geometry, precise field control, and a symmetric detector arrangement are all designed to suppress systematic effects that would mimic an EDM signal~\cite{Adelmann:muEDM2025}.

\subsection{Muonium spectroscopy}
\label{sec:muonium}

Muonium, the purely leptonic bound state of $\mu^+$ and $e^-$, is free from nuclear-structure corrections and therefore provides a clean system for testing bound-state QED and for determining fundamental constants such as the positive muon magnetic moment and the muon mass. Current programs \AFedit{combine} microwave spectroscopy of the hyperfine structure and \AFedit{Lamb-shift-related transitions with} laser spectroscopy of the $1S$--$2S$ transition.

J-PARC MuSEUM uses the intense pulsed muon beam at MUSE to produce muonium, drives the HFS transition with microwaves, and detects the resonance through decay positrons. MuSEUM observed the HFS resonance near zero magnetic field with $0.9~\mathrm{ppm}$ precision~\AFedit{\cite{Kanda:MuSEUM2020}}, and a subsequent Rabi-oscillation analysis of the same data reached a statistical uncertainty of $160~\mathrm{ppb}$~\cite{Nishimura:MuSEUM2021}. Future high-field measurements at the MUSE H-line, based on Rabi-oscillation spectroscopy, aim to improve on the current $12~\mathrm{ppb}$ high-field benchmark by about one order of magnitude~\cite{Kanda:MuSEUM2020}. The same apparatus concept can also be applied to the muonic helium HFS, which determines the negative muon magnetic moment and mass and tests three-body bound-state QED, with a projected precision gain of about a factor of one hundred~\cite{Strasser:MuSEUM2025}.

PSI Mu-MASS measures the muonium $1S$--$2S$ transition and aims for an uncertainty of about $10~\mathrm{kHz}$, three orders of magnitude below the $9.8~\mathrm{MHz}$ uncertainty of the current value. A keV-scale $\mu^+$ beam at the PSI LEM beamline is stopped in a mesoporous SiO$_2$ thin-film target, which emits muonium into vacuum. The atoms are excited to the $2S$ state by two-photon absorption in a $244~\mathrm{nm}$ continuous-wave laser field and are then ionized by a $355~\mathrm{nm}$ pulsed laser. The released $\mu^+$ is collected by an MCP detector, while surrounding scintillators tag the decay positron. Because the two-photon excitation probability is low, the measurement relies on a high $\mu^+$ rate and on stable ultraviolet laser operation, and will therefore benefit directly from the LEM beamline upgrades and from HIMB~\cite{Cortinovis:MuMASS2023}. Measurements at the same beamline have also determined the muonium Lamb shift and the $2S$ hyperfine splitting~\cite{Janka:Muonium2S2P2022}.

\subsection{Muonic atom spectroscopy}
\label{sec:muonic atoms}

In muonic atoms a negative muon occupies an atomic orbital roughly 200 times smaller than its electronic counterpart, so the energy levels are highly sensitive to the nuclear charge distribution, the Zemach radius, and nuclear polarization. \AFedit{Laser spectroscopy probes Lamb shifts and hyperfine splittings in muonic hydrogen and helium, while X-ray spectroscopy of muonic cascades covers a wider range of nuclei}. The 2010 muonic hydrogen Lamb shift measurement gave rise to the proton radius puzzle and motivated cross-checks among electronic, muonic, and scattering experiments~\cite{Pohl:ProtonRadius2010,Ohayon:MuonicXRays2023,Xiong:ProtonRadiusReview2023}.

The CREMA collaboration at PSI measured the $2S$--$2P$ Lamb shifts of muonic hydrogen, deuterium, and both helium isotopes, providing the most precise charge radii of these light nuclei. Its ongoing HyperMu experiment aims to measure the ground-state HFS of muonic hydrogen to about $1~\mathrm{ppm}$. A $6.8~\mathrm{\upmu m}$ pulsed laser drives the transition, and laser-excited atoms gain kinetic energy in collisions and are registered through the X-rays they induce at the target walls. A first resonance search is in preparation at the $\pi$E5 beamline and would give access to the proton Zemach radius~\cite{Amaro:HyperMu2022}.

FAMU measures the ground-state HFS of muonic hydrogen at the RIKEN-RAL muon facility at ISIS. A tunable mid-infrared laser near 6790~nm drives the transition in a cryogenic pressurized gas target. Laser-excited atoms transfer their muons to oxygen admixed in the gas, and the delayed oxygen X-rays are read out by fast LaBr$_3$:Ce scintillators and calibrated with HPGe detectors. The experiment has been taking data since 2023 and is scanning the laser frequency in search of the resonance. Its design goal is a determination of the proton Zemach radius to better than 1\%~\cite{Bonesini:FAMUDet2025,FAMU:Operation2025}.

\AFedit{At the J-PARC D2 beamline, a 240-pixel transition-edge sensor (TES) array was used to measure X-rays from muonic neon formed by stopping negative muons in a low-pressure gas target. The detector achieved an energy resolution of $5.2$--$5.5~\mathrm{eV}$ FWHM near $6.3~\mathrm{keV}$. The measured $5g$--$4f$ transition energies agreed with bound-state QED calculations, demonstrating a method for testing QED in strong electric fields~\cite{Okumura:MuonicNeon2023}.}

\AFedit{The QUARTET program at PSI aims to determine the charge radii of nuclei from lithium to neon using metallic magnetic calorimeters (MMCs)~\cite{Ohayon:MuonicXRays2023}. At the $\pi$E1 beamline, negative muons are stopped in solid targets, and the emitted X-rays are recorded by a 64-pixel MMC array. A recent measurement of muonic $^9$Be achieved an energy resolution of about $20~\mathrm{eV}$ FWHM near $33~\mathrm{keV}$. The $2p$--$1s$ transition energy was determined with an uncertainty of about $0.3~\mathrm{eV}$, improving the previous precision by a factor of 30 and enabling an updated determination of the nuclear charge radius~\cite{Eizenberg:MuonicBe2026}.}

For medium- and high-$Z$ nuclei, the muX experiment at PSI transfers muons stopped in a $100~\mathrm{bar}$ hydrogen--deuterium gas cell onto microgram-scale targets, extending muonic X-ray spectroscopy to rare and radioactive isotopes, with a first measurement of $^{248}$Cm and with the technique being developed toward $^{226}$Ra for the charge-radius determination needed in atomic parity violation experiments~\cite{Adamczak:muX2023}.

\begin{figure}[htbp]
\centering
\includegraphics[width=0.9\textwidth]{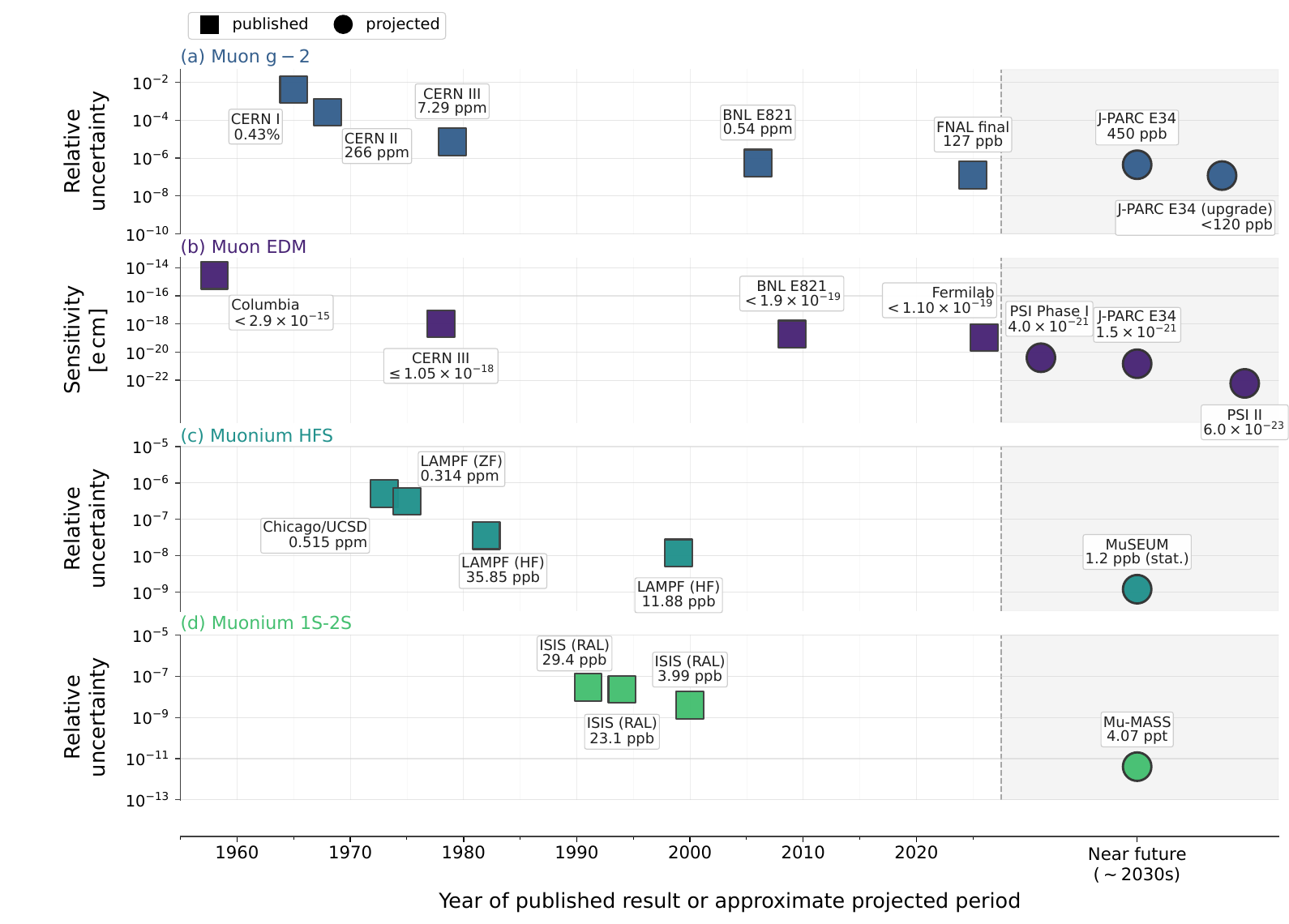}
\caption{Evolution of experimental precision for a selection of representative muon precision experiments:
         (a) the muon anomalous magnetic moment~\AFedit{\cite{Charpak:MuonG21965,Bailey:MuonG21968,Bailey:MuonG21979,Bennett:E8212006,Abi:MuonG22021,Aguillard:MuonG22023,MuonG2:Measurement2025,Abe:JPARCG2EDM2019,JPARC:g2EDMStatus2025}},
         (b) the muon EDM~\AFedit{\cite{Berley:MuonEDM1958,Bailey:MuonEDM1978,Bennett:MuonEDM2009,MuonG2:EDM2026,Adelmann:muEDM2025,Abe:JPARCG2EDM2019}},
         (c) the muonium ground-state HFS~\cite{Favart:MuoniumHFS1973,Casperson:MuoniumHFS1975,Mariam:MuoniumHFS1982,Liu:MuoniumHFS1999,Strasser:MuSEUM2025}, and
         (d) the muonium $1S$--$2S$ transition frequency~\cite{Jungmann:Muonium1S2S1991,Maas:Muonium1S2S1994,Meyer:Muonium1S2S2000,Cortinovis:MuMASS2023}.
         The plotted values combine published total uncertainties and projected targets; panel~(b) is shown as upper limits and projected sensitivities in units of $e\,$cm.}
\label{fig:precision-sensitivity}
\end{figure}

Figure~\ref{fig:precision-sensitivity} shows how the experimental precision of the dipole-moment and muonium observables has evolved, based on a selection of representative results for each. Muonic atom spectroscopy is not shown, as its results refer to different nuclei and observables rather than a single precision trend. Across all these programs, the projected improvements rely on the beam developments described in Section~\ref{sec:beams}, with requirements on intensity, time structure, and emittance that differ from one measurement to another.

\section{High-precision muon scattering experiments}
\label{sec:scattering}

Muon-scattering experiments \AFedit{use reconstructed tracks and kinematics to probe interactions between incident muons and targets over a broad momentum range}. \AFedit{This section discusses PSI MUSE, which uses the $\pi$M1 beamline for low-momentum elastic $e^{\pm}p$ and $\mu^{\pm}p$ scattering to compare lepton species and charge polarities in studies of lepton universality, the proton form factor and charge radius, and two-photon exchange~\cite{Gilman:MUSE2017,Cline:PiM1MUSE2022,Li:MUSERadiative2024,Lin:MUSECalorimeter2025};} MUonE, which uses high-energy elastic $\mu e$ scattering to determine the hadronic contribution to the running electromagnetic coupling in the space-like region and thereby the leading-order hadronic vacuum-polarization contribution to $a_\mu$~\cite{Abbiendi:MUonE2017,Abbiendi:MUonELOI2019,Gurgone:MUonETheory2024}; NA64$\mu$, which searches for light invisible particles produced in high-energy muon--nucleus interactions through missing momentum and missing downstream activity~\cite{Andreev:NA64muPRL2024,Andreev:NA64muPRD2024}; and PKMu, which investigates anomalous deflections from elastic muonphilic-dark-matter scattering, missing momentum in $\mu e^-\to\mu e^-X$, and resonant LFV $\mu^+e^-\to\phi$ annihilation. For the resonant channel, atomic binding and the target-electron momentum distribution affect the resonance and its signal prediction~\cite{Ruzi:PKMu2023,Gao:PKMuOverview2025,Liu:PKMuNewPhysics2026,Shen:LFVScalar2026}.

\subsection{\AFedit{Measuring low-momentum lepton--proton scattering with MUSE}}

\AFedit{MUSE at PSI's $\pi$M1 beamline measures elastic $e^\pm p$ and $\mu^\pm p$ cross sections at nominal momenta of $115$, $160$, and $210~\mathrm{MeV}/c$~\cite{Lin:MUSECalorimeter2025}, covering $Q^2\simeq0.002$--$0.08~\mathrm{GeV}^2$~\cite{Li:MUSERadiative2024}. Comparing electron with muon and positive with negative cross sections tests lepton universality and two-photon exchange, respectively~\cite{Gilman:MUSE2017,Lin:MUSECalorimeter2025}. The design report targets proton-radius precision better than $0.01~\mathrm{fm}$~\cite{Gilman:MUSE2017}. A recent dispersive chiral-EFT study identifies $210~\mathrm{MeV}/c$ and $Q^2\simeq0.05$--$0.08~\mathrm{GeV}^2$ as favorable for radius extraction, requiring about $0.5\%$ cross-section precision for $0.01~\mathrm{fm}$ radius accuracy~\cite{GilDominguez:MUSERadius2023}.}

\AFedit{The mixed secondary beam contains electrons, muons, and pions; its momentum dispersion, beam spot, and timing have been measured and compared with transport simulations~\cite{Cline:PiM1MUSE2022}. Timing relative to the accelerator RF identifies beam species, while dedicated time-of-flight measurements determine beam momentum~\cite{Cline:PiM1MUSE2022,Rostomyan:MUSEtiming2021}; either beam polarity can be selected~\cite{Lin:MUSECalorimeter2025}. The cryogenic liquid-hydrogen target is designed for stable density and minimal surrounding material; its ladder also holds an empty cell, polyethylene and carbon targets for vertex reconstruction and alignment, and a beam-focus monitor~\cite{Roy:MUSEHydrogen2020}. Two symmetric spectrometers cover $20^\circ$--$100^\circ$ and provide simultaneous cross-section measurements as a check on systematics~\cite{Gilman:MUSE2017}. Straw chambers and scintillator planes track and time scattered particles, while upstream GEM chambers measure incident trajectories and the beam hodoscope provides particle identification and the start time for time-of-flight measurements~\cite{Gilman:MUSE2017,Rostomyan:MUSEtiming2021}.}

\AFedit{The precision goals require control of radiative effects and backgrounds~\cite{Gilman:MUSE2017,Li:MUSERadiative2024}. The radiative-correction study estimates instrumental uncertainties, including the effect of hard initial-state-radiation events detected in the calorimeter and excluded from the cross-section sample~\cite{Li:MUSERadiative2024}. A beamline lead-glass calorimeter measures incident energy and tags forward photons relevant to these corrections~\cite{Lin:MUSECalorimeter2025}. Pion scattering and beam-tail particles striking chamber support posts produce trigger backgrounds~\cite{Gilman:MUSE2017,Ratvasky:MUSEPostVeto2026}. The trigger combines beam identification, scattered-particle detection, and veto information; a target-chamber post-veto detector suppresses support-post backgrounds at trigger level~\cite{Gilman:MUSE2017,Ratvasky:MUSEPostVeto2026}.}

\subsection{Determining hadronic vacuum polarization with MUonE}

The leading-order hadronic vacuum-polarization contribution is the dominant source of uncertainty in the Standard-Model prediction of $a_\mu$~\cite{Keshavarzi:MuonG2Review2021,Gurgone:MUonETheory2024}. MUonE measures the shape of the elastic $\mu e\to\mu e$ differential cross section to extract the hadronic running of $\alpha$ at space-like momentum transfer, using the low-$q^2$ region for an internal normalization~\cite{Abbiendi:MUonE2017,Abbiendi:MUonELOI2019,MUonE:PhaseI2024}. The full measurement aims for a statistical uncertainty of about $0.3\%$ on this contribution after two years of data taking~\cite{Abbiendi:MUonE2017}; reaching that precision requires control of the differential cross section at the $10~\mathrm{ppm}$ level~\cite{Abbiendi:MUonE2017,Gurgone:MUonETheory2024}.

The experiment uses the high-intensity $160~\mathrm{GeV}$ muon beam at the CERN SPS M2 line~\cite{Abbiendi:MUonELOI2019,MUonE:PhaseI2024}. The baseline full detector comprises $40$ target--tracker stations; each station combines a $1.5~\mathrm{cm}$ beryllium target with three pairs of silicon strip planes to reconstruct the correlated electron and muon scattering angles~\cite{Abbiendi:MUonELOI2019}. The three-station Phase-I demonstrator uses six CMS 2S silicon-strip modules per station, arranged in three pairs; the outer pairs are tilted by about $233~\mathrm{mrad}$ to obtain a spatial resolution better than about $15~\mathrm{\upmu m}$, and the Invar support frames provide mechanical stability below $10~\mathrm{\upmu m}$~\cite{Pesaresi:MUonEPhaseI2026}. Its $5\times5$ PbWO$_4$ electromagnetic calorimeter is designed for an energy resolution of about $1\%$ above $100~\mathrm{GeV}$ and an impact-position resolution below $1~\mathrm{mm}$~\cite{Pesaresi:MUonEPhaseI2026}. Together with the muon filter and prototype beam-momentum spectrometer, this apparatus achieved stable $40~\mathrm{MHz}$ operation, sub-nanosecond synchronization, and recorded more than $5\times10^{11}$ events in the 2025 run~\cite{Pesaresi:MUonEPhaseI2026}. The ongoing Phase-I analysis aims for a first measurement with a projected $20\%$ statistical uncertainty~\cite{MUonE:PhaseI2024}.

The station array provides the large elastic-event sample needed to measure the differential-cross-section shape with the required precision~\cite{Abbiendi:MUonE2017,Abbiendi:MUonELOI2019}. Its principal experimental challenges are control of multiple scattering and alignment, accurate reconstruction of the incident muon momentum, and a detector simulation that incorporates radiative and lepton-pair backgrounds~\cite{Abbiendi:MUonETestBeam2021,MUonE:PhaseI2024,Gurgone:MUonETheory2024}. The Phase-I measurements are therefore both an initial physics measurement and a validation of the experimental inputs required for the full HVP determination~\cite{MUonE:PhaseI2024,Pesaresi:MUonEPhaseI2026}.

\subsection[Searching for invisible dark sectors with \texorpdfstring{NA64$\mu$}{NA64mu}]{Searching for invisible dark sectors with \texorpdfstring{NA64$\boldsymbol{\mu}$}{NA64mu}}

Light invisible particles coupled to muons are motivated by the muon $g-2$ discrepancy and by thermal-dark-matter scenarios; in a $U(1)_{L_\mu-L_\tau}$ model, a $Z^\prime$ can address both~\cite{MuonG2:Measurement2025,Keshavarzi:MuonG2Review2021,Holst:MuonG2ThermalDM2022}. NA64$\mu$ tests such models through the bremsstrahlung-like reaction $\mu N\to\mu N X$, followed by an invisible decay of $X$. A signal is a single scattered muon carrying less than about half of the incident beam momentum together with no additional visible activity in the downstream detectors~\cite{Andreev:NA64muPRL2024,Andreev:NA64muPRD2024}.

The experiment uses the $160~\mathrm{GeV}/c$ CERN SPS M2 muon beam, the same beamline used by MUonE. Its upstream magnetic spectrometer combines three $5~\mathrm{T\,m}$ dipoles with four Micromegas detectors, two straw chambers, and six scintillator hodoscopes, yielding an incoming-momentum resolution of about $3.8\%$~\cite{Andreev:NA64muPRL2024}. The active electromagnetic-calorimeter target is $40$ radiation lengths deep and is followed by a $55\times55~\mathrm{cm}^2$ veto counter and a $5$ interaction-length hadronic calorimeter. A downstream spectrometer with a $1.4~\mathrm{T\,m}$ dipole, four GEM trackers, two straw chambers, and three Micromegas detectors reconstructs the scattered-muon momentum with about $4.4\%$ resolution; two \(120\times60~\mathrm{cm}^2\) hadronic calorimeter modules, each approximately 15 nuclear interaction lengths thick, provide the final hermetic veto. In its 2022 data set of $(1.98\pm0.02)\times10^{10}$ muons on target, NA64$\mu$ observed no events in the signal region and set new constraints on the remaining $L_\mu-L_\tau$ parameter space that could explain the muon anomalous magnetic moment. The same technique probes invisibly decaying scalar and vector mediators, including dark-photon scenarios, and complements NA64 searches with electron and positron beams~\cite{Andreev:NA64muPRD2024}.

The reach of this program is set by the accumulated muons on target together with the efficiency and rejection of the tracking, veto, and calorimeter systems. For the post-LS3 Run~4 programme, NA64$\mu$ plans an upgraded M2 setup and an exposure of $2\times10^{13}$ muons on target~\cite{NA64:Status2025}. Background predictions are tested with dedicated control samples before the missing-momentum signal region is unblinded, making detector-response validation as important as additional beam exposure~\cite{Andreev:NA64muPRL2024,Andreev:NA64muPRD2024}. This experimental strategy permits a common missing-momentum apparatus to test several muonphilic mediator hypotheses without changing the underlying observable~\cite{Andreev:NA64muPRD2024}.

\subsection{Setting direct limits on muon--dark-matter scattering with PKMu}

Dark matter that couples preferentially to muons is complementary to scenarios tested through conventional nuclear- or electron-recoil searches. PKMu uses the muon as the projectile: rather than waiting for a dark-matter particle to scatter from detector matter, it searches for the rare scattering of an incident muon from dark matter through an anomalous track deflection~\cite{Ruzi:PKMu2023,Liu:PKMuNewPhysics2026}. PKMu follows a phased program: Phase~I uses cosmic-ray muons to establish the detector, reconstruction, and background-control methods, whereas Phase~II transfers the search to a dedicated accelerator beam with controlled exposure and detector conditions~\cite{Liu:PKMuCosmicDM2026,Zhang:MuonDMBeam2026}.

The angular-scattering system uses either a four- or a six-layer gaseous tracker to reconstruct the incident and outgoing trajectories and their point of closest approach (PoCA). Glass RPCs with delay-line readout provide large-area, two-dimensional measurements, while GEMs are a complementary option where higher local rate capability or finer segmentation is needed~\cite{Yu:PKMuTomography2024,Sauli:GEM1997,Riegler:RPC2004}. The segment separation, active area, position resolution, and relative alignment set the kink sensitivity; in-situ calibration with through-going cosmic-ray tracks is therefore essential~\cite{Yang:DetectorPositionCalibration2026}.

The Phase-I $63~\mathrm{day}$ cosmic-ray measurement set a 95\% CL limit of $1.61\times10^{-17}~\mathrm{cm}^2$ at $m_\chi=1~\mathrm{GeV}$, assuming an Earth-bound density enhancement~\cite{Liu:PKMuCosmicDM2026}. For Phase~II, a Geant4 study of the four-layer PKMu RPC geometry ($28\times28~\mathrm{cm}^2$ active area per layer) uses the reconstructed track kink as the signal observable~\cite{Zhang:MuonDMBeam2026}. With a realistic HIAF-like $\sim3.3$--$3.5~\mathrm{GeV}$ beam at $10^5~\mu/\mathrm{s}$, a one-day exposure is projected to improve the Phase-I limit by nearly two orders of magnitude. Provided that the beam is contained in the active area, the reach is governed mainly by the total muon intensity, effective fiducial path length, and angular resolution; extending the tracking lever arm is more effective than changing the transverse beam size~\cite{Zhang:MuonDMBeam2026}.

Complementary reconstruction methods can improve both the dark-matter search and other precision scattering measurements. Projection-shifted MUon transMission tomogrAphy (P$\mu$MA) connects upstream and downstream hits and maps a material-induced deflection to a displacement in a projected imaging plane. It permits fewer tracking planes and hence greater acceptance, while a multi-plane layout can recover scattering-angle information for material discrimination~\cite{Qin:ProjectionShiftedMuonTomography2025}. A \AFedit{residual-likelihood reconstruction uses downstream hit residuals, marginalizes the unmeasured muon momentum, and incorporates open-field calibration to reconstruct the material distribution with a magnet-free tracker}~\cite{Zhao:RawHitMuonTomography2026}. Together, these approaches can increase usable statistics and improve spatial and angular resolution, strengthening signal--background discrimination.

\subsection{Searching for light dark sectors with PKMu}

NA64$\mu$ uses missing momentum in $160~\mathrm{GeV}$ muon--nucleus interactions to set limits on invisibly decaying muonphilic mediators, but its sensitivity is limited in the low-mass region near $m_{Z^\prime}\sim10~\mathrm{MeV}$~\cite{Andreev:NA64muPRL2024,Andreev:NA64muPRD2024}. To complement this low-mass reach, the proposed HIAF--PKMu search is designed to identify the scattered muon and electron in GeV-scale $\mu e^-\to\mu e^-X$ events~\cite{Wang:LightDarkSectors2026}. For a $10~\mathrm{GeV}$ muon beam and a $20~\mathrm{mm}$ graphite target, the single-station HIAF--PKMu proposal projects a 95\% CL sensitivity to $g_{Z^\prime}$ at the $10^{-3}$ level, with its strongest reach for $m_{Z^\prime}$ near $10~\mathrm{MeV}$~\cite{Wang:LightDarkSectors2026}.

Drawing on the MUonE detector design for precision two-dimensional lepton tracking, the proposed single station places silicon-strip modules downstream of an electron-rich target~\cite{Wang:LightDarkSectors2026,Abbiendi:MUonELOI2019}. Each module provides orthogonal $x$--$y$ strip measurements, while additional $45^\circ$ and $135^\circ$ planes provide redundant projections for three-dimensional pattern recognition and the resolution of track ambiguities~\cite{Wang:LightDarkSectors2026,MUonE:ECAL2024}. A scintillator-based particle-identification system, optionally supplemented by an electromagnetic calorimeter, distinguishes the scattered muon and electron and measures the electron energy. A boosted decision tree based on reconstructed kinematic variables suppresses Standard-Model backgrounds~\cite{Wang:LightDarkSectors2026}.

The same strategy benefits strongly from a multi-station implementation. In a 40-station MUonE-style layout with a $10~\mathrm{cm}$ lead target at each station, the larger target thickness and electron density increase the expected yield by almost three orders of magnitude. The projected reach would surpass current constraints near $m_{Z^\prime}\sim10~\mathrm{MeV}$, and remains competitive for a conservative one-year exposure of $10^{13}$ muons on target~\cite{Wang:LightDarkSectors2026}. This gain makes the control of beam purity, tracking efficiency, and material-induced backgrounds central to a future high-statistics measurement.

\subsection{Probing LFV boson resonances and atomic binding effects with PKMu}

An $e$--$\mu$ flavor-changing scalar would provide a direct signal of physics beyond the Standard Model. Positive muon beams can test this possibility through the $s$-channel reaction $\mu^+e^-\to\phi$, in which resonant production can greatly enhance sensitivity to the LFV coupling; an invisibly decaying $\phi$ leaves a single incoming muon pointing to the target with no downstream charged particle. This channel also provides a direct probe of bound-electron effects: the target-electron binding energies and momentum distribution determine the target-dependent resonance shape and rate. Measuring this distortion is therefore a physics objective in its own right, as well as essential to a reliable LFV-signal prediction~\cite{Shen:LFVScalar2026}. At the anticipated HIAF positive-muon intensity, the compact system is projected to test $e$--$\mu$ LFV couplings at the $10^{-5}$ level near resonance in less than $1~\mathrm{day}$ of running~\cite{Shen:LFVScalar2026}.

The detector geometry remains to be optimized, but the essential requirements are clear: the incident muon track must be reconstructed and the target region must be monitored for charged-particle activity. Multiple RPC and scintillator elements can be combined into a redundant downstream tagging system, with both detector technologies contributing to the veto and its calibration.

The target and veto choices set the dominant systematics. The target material must balance a large atomic-electron density against ordinary electromagnetic activity and the corresponding bound-electron calculation. At a realistic per-plane RPC efficiency near 95\%, a miss in an individual element is not rare, so the rejection and calibration must use the joint response of the full detector system. Signal and control regions can be defined from the combined RPC and scintillator responses, and their yields fitted simultaneously to determine tagging efficiencies and correlations in situ. This constrains charged-particle leakage into the signal region and propagates the corresponding uncertainty to the signal yield.

\begin{figure}[htbp]
\centering
\includegraphics[width=0.8\textwidth]{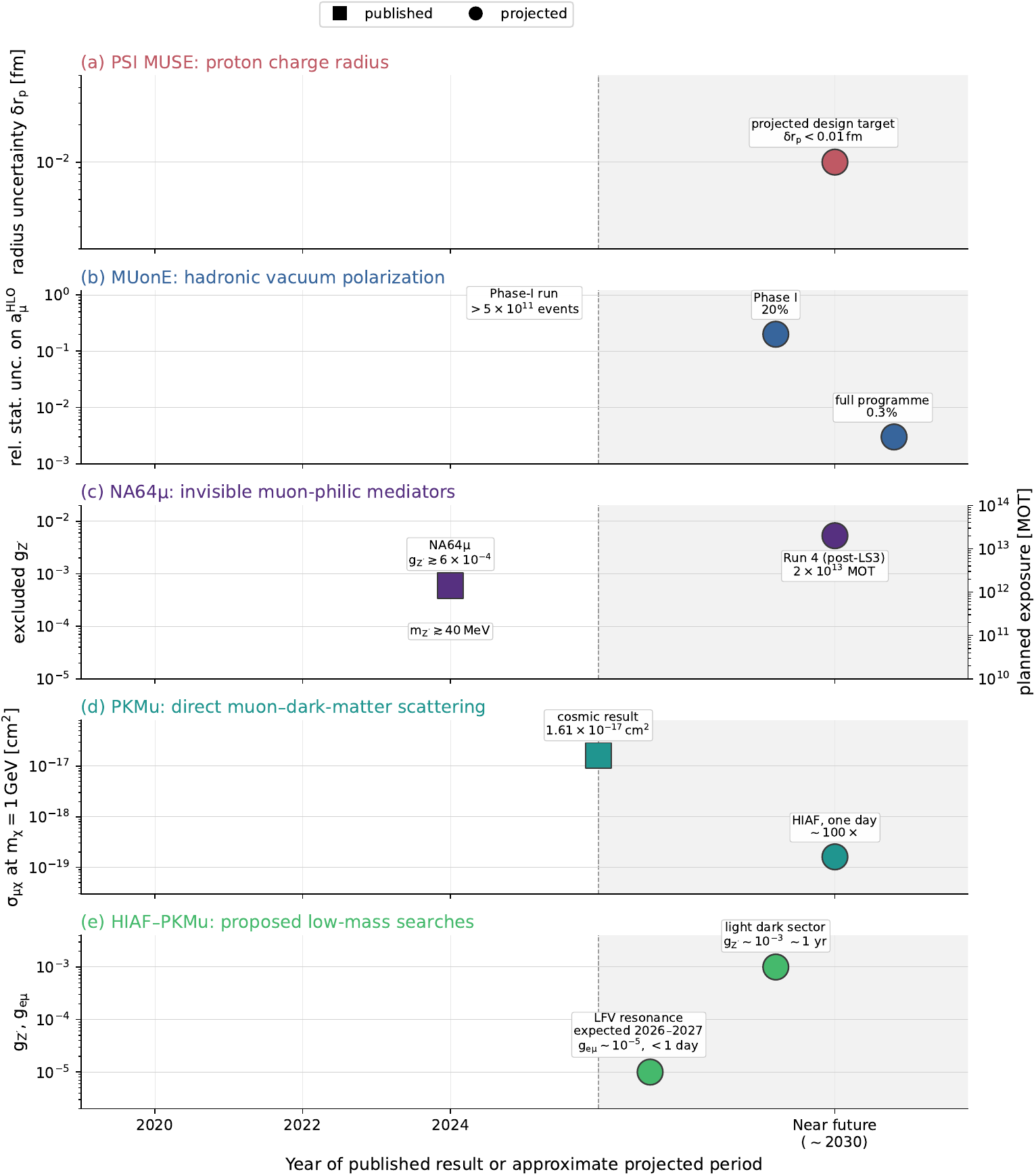}
\caption{Status of selected muon-scattering programmes: (a) \AFedit{PSI MUSE projected proton-radius design sensitivity}, (b) MUonE determination of hadronic vacuum polarization, (c) NA64$\mu$ searches for invisible muonphilic mediators, (d) PKMu direct muon--dark-matter scattering, and (e) proposed HIAF--PKMu light-dark-sector and LFV searches. \AFedit{Square markers denote published results or completed data taking, whereas circular markers denote projected targets; the MUSE marker is a design target, not a measured radius, and its near-future placement is illustrative rather than a forecasted publication date}. In (c), the right axis gives the planned Run~4 exposure; all other axes are observable-specific and are not compared directly~\AFedit{\cite{Abbiendi:MUonE2017,MUonE:PhaseI2024,Pesaresi:MUonEPhaseI2026,Andreev:NA64muPRL2024,NA64:Status2025,Liu:PKMuCosmicDM2026,Zhang:MuonDMBeam2026,Wang:LightDarkSectors2026,Shen:LFVScalar2026,Gilman:MUSE2017,GilDominguez:MUSERadius2023}}.}
\label{fig:scattering-status}
\end{figure}

\AFedit{Figure~\ref{fig:scattering-status} summarizes selected published results and projected targets across the scattering programs. Only the proton-radius benchmark is plotted for PSI MUSE; its lepton-universality and two-photon-exchange goals are described in the subsection above~\cite{Gilman:MUSE2017,GilDominguez:MUSERadius2023}}. The NA64$\mu$ exposure is displayed separately because its translation into coupling reach depends on the mediator model and mass~\AFedit{\cite{Abbiendi:MUonE2017,MUonE:PhaseI2024,Pesaresi:MUonEPhaseI2026,Andreev:NA64muPRL2024,NA64:Status2025,Liu:PKMuCosmicDM2026,Zhang:MuonDMBeam2026,Wang:LightDarkSectors2026,Shen:LFVScalar2026}}.

\section{Summary and outlook}
\label{sec:summary}

The future experimental program can be organized into three broad categories. CLFV searches ask whether charged-lepton flavor changes: Mu2e and COMET Phase~II target $\mathcal{O}(10^{-17})$ sensitivity in $\mu^-N\to e^-N$ conversion, Mu3e Phase~II targets $\mathcal{O}(10^{-16})$ in $\mu^+\to e^+e^-e^+$, MEG~II has reached the $10^{-13}$ level in $\mu^+\to e^+\gamma$, and MACE will search for muonium--antimuonium conversion with a multi-coincidence detector. Precision experiments instead determine flavor-conserving muon, muonium, and muonic-atom observables: Fermilab Muon $g-2$ has reached $127~\mathrm{ppb}$ on $a_\mu$, with a combined Fermilab--BNL average of $124~\mathrm{ppb}$; J-PARC $g-2$/EDM and PSI muEDM provide complementary dipole-moment measurements; and MuSEUM, Mu-MASS, and FAMU advance tests of bound-state QED and determinations of nuclear properties such as the proton Zemach radius. \AFedit{Scattering experiments cover complementary kinematic regimes: PSI MUSE compares low-momentum elastic $\mu p$ and $ep$ scattering at $115$--$210~\mathrm{MeV}/c$ to study the proton radius, lepton universality, and two-photon exchange; MUonE determines hadronic vacuum polarization; NA64$\mu$ searches for invisible dark sectors; and PKMu investigates muonphilic-dark-matter scattering, light mediators, LFV interactions, and bound-electron effects in resonant $\mu^+e^-$ annihilation~\cite{Gilman:MUSE2017,Abbiendi:MUonELOI2019,Andreev:NA64muPRL2024,Andreev:NA64muPRD2024,Liu:PKMuNewPhysics2026,Shen:LFVScalar2026}}.

Further progress depends on the coupled development of muon beam facilities and experimental apparatus. On the beam side, upgrades and new facilities including PSI HIMB, Fermilab PIP-II, and Chinese projects such as MELODY, CiADS, and HIAF are expanding the accessible parameter space in intensity, time structure, emittance, and particle purity. Advanced muon beam techniques (slow, ultra-slow, reaccelerated, and phase-rotated beams) further improve phase-space control and background conditions. On the detector side, CLFV searches are moving toward ultra-thin tracking detectors, high-granularity calorimeters, and fast coincidence systems to cope with accidental backgrounds at higher rates. Precision measurements require low-emittance injection, highly uniform storage fields, frozen-spin traps, stable laser systems, \AFedit{high-resolution X-ray microcalorimeters} and fast X-ray detectors. \AFedit{Muon-scattering programs require apparatus matched to their kinematics: MUSE combines a liquid-hydrogen target with straw-tube and GEM tracking and time-of-flight identification, while MUonE requires precision lepton tracking and missing-momentum searches require momentum measurement and calibrated veto systems~\cite{Gilman:MUSE2017,Abbiendi:MUonELOI2019,Andreev:NA64muPRL2024}}. The combination of beam upgrades and detector optimization is expected to improve sensitivities or precision by one to several orders of magnitude relative to present benchmarks.

Looking ahead, each category has a distinct limiting requirement. For CLFV, $\mu^-N\to e^-N$ conversion offers favorable scalability at high-intensity negative-muon beams because it has lower accidental-coincidence backgrounds; PRISM/AMF-type phase-space purification could support experiments beyond Mu2e-II. Positive-muon rare-decay channels such as $\mu^+\to e^+\gamma$ require new detector concepts to control rate-dependent backgrounds. For precision observables, the final Fermilab Muon $g-2$ result remains a key reference, while the compact J-PARC storage magnet provides an independent $a_\mu$ test and the PSI frozen-spin trap targets the muon EDM. A possible Fermilab follow-up using the existing storage ring and PIP-II beam capability aims for a precision near $40~\mathrm{ppb}$~\cite{Hertzog:Hoferichter2026}, and muonium and muonic-atom spectroscopy will rely on higher statistics, more stable lasers, and faster detectors. For scattering, \AFedit{MUSE requires control of mixed-beam particle identification and charge-dependent effects in its lepton--proton comparisons;} MUonE requires ppm-level control of tracking, alignment, beam momentum, and theory; NA64$\mu$ requires efficient momentum reconstruction and vetoing; and PKMu requires a stable angular lever arm, target and tracker designs matched to the signal, and calibrated multi-detector vetoes for invisible final states~\cite{Gilman:MUSE2017,Gurgone:MUonETheory2024,Andreev:NA64muPRL2024,Zhang:MuonDMBeam2026,Shen:LFVScalar2026}. Continued advances in beams and apparatus will strengthen all three categories as well as their explicitly identified overlaps.

\acknowledgments
The authors would like to acknowledge the MACE Working Group for their invaluable discussion. This work was supported in part by the National Natural Science Foundation of China under Grant Nos.~12325504, 12347105 and 12075326, Guangdong Basic and Applied Basic Research Foundation under Grant No.~2025A1515010669, the Natural Science Foundation of Guangzhou under Grant No.~2024A04J6243. This research was supported by the Munich Institute for Astro-, Particle and BioPhysics (MIAPbP) which is funded by the Deutsche Forschungsgemeinschaft (DFG, German Research Foundation) under Germany's Excellence Strategy--EXC-2094--390783311.


\bibliographystyle{JHEP}
\bibliography{references}

@article{Keshavarzi:MuonG2Review2021,
    archiveprefix = {arXiv},
    author = {Alex Keshavarzi and Kim Siang Khaw and Tamaki Yoshioka},
    doi = {10.1016/j.nuclphysb.2022.115675},
    eprint = {2106.06723},
    journal = {Nucl. Phys. B},
    pages = {115675},
    primaryclass = {hep-ex},
    title = {Muon $g-2$: A review},
    volume = {975},
    year = {2022}
}

@article{Hertzog:Hoferichter2026,
    archiveprefix = {arXiv},
    author = {David W. Hertzog and Martin Hoferichter},
    doi = {10.1146/annurev-nucl-102422-040841},
    eprint = {2512.16980},
    journal = {Annu. Rev. Nucl. Part. Sci.},
    pages = {343--368},
    primaryclass = {hep-ph},
    title = {The anomalous magnetic moment of the muon: status and perspectives},
    volume = {76},
    year = {2026}
}

@article{PDG2024,
    author = {{Particle Data Group}},
    doi = {10.1103/PhysRevD.110.030001},
    journal = {Phys. Rev. D},
    pages = {030001},
    title = {Review of Particle Physics},
    volume = {110},
    year = {2024}
}

@article{Bai:SnowmassMACE2022,
    archiveprefix = {arXiv},
    author = {Ai-Yu Bai and Yu Chen and Yukai Chen and Rui-Rui Fan and Zhilong Hou and Han-Tao Jing and Hai-Bo Li and others},
    eprint = {2203.11406},
    primaryclass = {hep-ph},
    title = {Snowmass2021 Whitepaper: Muonium to antimuonium conversion},
    year = {2022}
}

@article{Bai:MACE2024,
    archiveprefix = {arXiv},
    author = {Ai-Yu Bai and Hanjie Cai and Chang-Lin Chen and others},
    doi = {10.1007/s41365-025-01876-0},
    eprint = {2410.18817},
    journal = {Nuclear Science and Techniques},
    pages = {57},
    primaryclass = {hep-ex},
    title = {Conceptual Design of the Muonium-to-Antimuonium Conversion Experiment ({MACE})},
    volume = {37},
    year = {2026}
}

@article{Bernstein:Mu2e2019,
    author = {R. H. Bernstein},
    doi = {10.3389/fphy.2019.00001},
    journal = {Front. Phys.},
    pages = {1},
    title = {The {Mu2e} Experiment},
    volume = {7},
    year = {2019}
}

@article{Kuno:COMET2013,
    author = {Y. Kuno},
    doi = {10.1093/ptep/pts089},
    journal = {Prog. Theor. Exp. Phys.},
    pages = {022C01},
    title = {A search for muon-to-electron conversion at {J-PARC}: the {COMET} experiment},
    volume = {2013},
    year = {2013}
}

@article{Mu3e:TDR2020,
    archiveprefix = {arXiv},
    author = {K. Arndt and H. Augustin and P. Baesso and others},
    doi = {10.1016/j.nima.2021.165679},
    eprint = {2009.11690},
    journal = {Nucl. Instrum. Methods Phys. Res. A},
    pages = {165679},
    primaryclass = {physics.ins-det},
    title = {Technical design of the {Phase I} {Mu3e} experiment},
    volume = {1014},
    year = {2021}
}

@article{MEGII:Result2025,
    author = {K. Afanaciev and others},
    doi = {10.1140/epjc/s10052-025-14906-3},
    journal = {Eur. Phys. J. C},
    pages = {1177},
    title = {New limit on the $\mu^+\to e^+\gamma$ decay with the {MEG II} experiment},
    volume = {85},
    year = {2025}
}

@article{Gilman:MUSE2017,
    archiveprefix = {arXiv},
    author = {Richard Gilman and Elizabeth J. Downie and Gerald Ron and Sebastian Strauch and others},
    doi = {10.48550/arXiv.1709.09753},
    eprint = {1709.09753},
    primaryclass = {physics.ins-det},
    title = {Technical Design Report for the {Paul Scherrer Institute} Experiment {R-12-01.1}: Studying the Proton ``Radius'' Puzzle with $\mu p$ Elastic Scattering},
    year = {2017}
}

@techreport{Abbiendi:MUonELOI2019,
    address = {Geneva},
    author = {G. Abbiendi and others},
    institution = {CERN},
    number = {CERN-SPSC-2019-026, SPSC-I-252},
    title = {Letter of Intent: the {MUonE} project},
    year = {2019}
}

@article{Andreev:NA64muPRL2024,
    author = {Y. M. Andreev and others},
    doi = {10.1103/PhysRevLett.132.211803},
    journal = {Phys. Rev. Lett.},
    pages = {211803},
    title = {First Results in the Search for Dark Sectors at {NA64} with the {CERN} {SPS} High Energy Muon Beam},
    volume = {132},
    year = {2024}
}

@article{Andreev:NA64muPRD2024,
    author = {Y. M. Andreev and others},
    doi = {10.1103/PhysRevD.110.112015},
    journal = {Phys. Rev. D},
    pages = {112015},
    title = {Shedding light on dark sectors with high-energy muons at the {NA64} experiment at the {CERN} {SPS}},
    volume = {110},
    year = {2024}
}

@article{Liu:PKMuNewPhysics2026,
    author = {Cheng-En Liu and Leyun Gao and Zijian Wang and others},
    doi = {10.1360/CSB-2025-5452},
    journal = {Chinese Science Bulletin},
    note = {In Chinese},
    number = {4},
    pages = {894--903},
    title = {Probing and knocking with muons and new physics exploration},
    volume = {71},
    year = {2026}
}

@article{Shen:LFVScalar2026,
    archiveprefix = {arXiv},
    author = {Jinhong Shen and Youpeng Wu and Zijian Wang and others},
    eprint = {2607.18669},
    primaryclass = {hep-ph},
    title = {Production of lepton-flavor-violating scalars through resonant positive-muon annihilation on atomic electrons},
    year = {2026}
}

@article{Prokscha:MuE4PSI2008,
    author = {T. Prokscha and E. Morenzoni and K. Deiters and others},
    doi = {10.1016/j.nima.2008.07.081},
    journal = {Nucl. Instrum. Methods Phys. Res. A},
    pages = {317--331},
    title = {The new {$\mu$E4} beam at {PSI}: A hybrid-type large acceptance channel for the generation of a high intensity surface-muon beam},
    volume = {595},
    year = {2008}
}

@techreport{Bartoszek:Mu2eTDR2015,
    address = {Batavia, Illinois},
    archiveprefix = {arXiv},
    author = {L. Bartoszek and E. Barnes and J. P. Miller and others},
    doi = {10.2172/1172555},
    eprint = {1501.05241},
    institution = {Fermi National Accelerator Laboratory},
    number = {FERMILAB-DESIGN-2014-01},
    primaryclass = {physics.ins-det},
    title = {{Mu2e} Technical Design Report},
    year = {2014}
}

@article{COMET:PhaseITDR2020,
    archiveprefix = {arXiv},
    author = {{COMET Collaboration}},
    doi = {10.1093/ptep/ptz125},
    eprint = {1812.09018},
    journal = {Prog. Theor. Exp. Phys.},
    pages = {033C01},
    primaryclass = {physics.ins-det},
    title = {{COMET} {Phase-I} Technical Design Report},
    volume = {2020},
    year = {2020}
}

@article{Stratakis:MuonCampus2017,
    author = {D. Stratakis and M. E. Convery and C. Johnstone and others},
    doi = {10.1103/PhysRevAccelBeams.20.111003},
    journal = {Phys. Rev. Accel. Beams},
    pages = {111003},
    title = {Accelerator performance analysis of the {Fermilab} {Muon Campus}},
    volume = {20},
    year = {2017}
}

@article{Shimomura:JPARCMUSE2024,
    author = {K. Shimomura and A. Koda and A. D. Pant and others},
    doi = {10.1007/s10751-024-01863-8},
    journal = {Interactions},
    pages = {31},
    title = {Pulsed muon facility of {J-PARC} {MUSE}},
    volume = {245},
    year = {2024}
}

@article{Xu:HIAFMuon2025,
    archiveprefix = {arXiv},
    author = {Yu Xu and Xueheng Zhang and Yuhong Yu and others},
    doi = {10.1103/PhysRevAccelBeams.28.053401},
    eprint = {2502.20915},
    journal = {Phys. Rev. Accel. Beams},
    pages = {053401},
    primaryclass = {physics.acc-ph},
    title = {Feasibility study of the {GeV}-energy muon source based on the High Intensity Heavy-Ion Accelerator Facility},
    volume = {28},
    year = {2025}
}

@article{Wang:LightDarkSectors2026,
    archiveprefix = {arXiv},
    author = {Zijian Wang and Leyun Gao and Zhuo Chen and others},
    doi = {10.1103/9srh-sw94},
    eprint = {2511.08950},
    journal = {Phys. Rev. D},
    pages = {072008},
    primaryclass = {hep-ex},
    title = {Search for light dark sectors with {GeV} muon beams},
    volume = {113},
    year = {2026}
}

@misc{PSI:CHRISP,
    author = {{Paul Scherrer Institute}},
    howpublished = {\url{https://www.psi.ch/en/ltp/facilities}},
    note = {Accessed: 29 October 2026},
    title = {{CHRISP -- Swiss Research Infrastructure for Particle Physics}}
}

@misc{TRIUMF:MuonBeamlines,
    author = {{TRIUMF Centre for Molecular and Materials Science}},
    howpublished = {\url{https://triumf.ca/research/cmms/}},
    note = {Accessed: 2026-11-06},
    title = {{Centre for Molecular and Materials Science}},
    year = {2026}
}

@article{Hillier:ISIS2019,
    author = {A. D. Hillier and J. S. Lord and K. Ishida and C. Rogers},
    doi = {10.1098/rsta.2018.0064},
    journal = {Philos. Trans. R. Soc. A},
    pages = {20180064},
    title = {Muons at {ISIS}},
    volume = {377},
    year = {2019}
}

@article{Miyake:JPARCMUSE2012,
    author = {Y. Miyake and K. Shimomura and N. Kawamura and others},
    doi = {10.1016/j.phpro.2012.04.037},
    journal = {Physics Procedia},
    pages = {46--49},
    title = {{J-PARC} Muon Facility, {MUSE}},
    volume = {30},
    year = {2012}
}

@article{Strasser:MuSEUM2025,
    archiveprefix = {arXiv},
    author = {Patrick Strasser and Mitsushi Abe and Kanta Asai and others},
    doi = {10.1140/epjd/s10053-025-00959-2},
    eprint = {2501.02736},
    journal = {Eur. Phys. J. D},
    pages = {20},
    primaryclass = {physics.atom-ph},
    title = {Precision measurements of muonium and muonic helium hyperfine structure at {J-PARC}},
    volume = {79},
    year = {2025}
}

@inproceedings{Nagaslaev:Mu2eSlowExtraction2026,
    author = {V. Nagaslaev and others},
    booktitle = {Proc. IPAC'26},
    doi = {10.18429/JACoW-IPAC2026-THP4310},
    title = {Slow Extraction Beam Commissioning for the {Mu2e} Experiment at {Fermilab}},
    year = {2026}
}

@techreport{Adams:AMBERProposal2019,
    address = {Geneva},
    author = {B. Adams and others},
    institution = {CERN},
    number = {CERN-SPSC-2019-022, SPSC-P-360},
    title = {Proposal for Measurements at the {M2} beam line of the {CERN} {SPS}},
    year = {2019}
}

@article{Abbiendi:MUonE2017,
    author = {G. Abbiendi and Carloni Calame, C. M. and U. Marconi and others},
    doi = {10.1140/epjc/s10052-017-4633-z},
    journal = {Eur. Phys. J. C},
    pages = {139},
    title = {Measuring the leading hadronic contribution to the muon $g-2$ via $\mu e$ scattering},
    volume = {77},
    year = {2017}
}

@article{Cook:MuSIC2017,
    archiveprefix = {arXiv},
    author = {S. Cook and R. D'Arcy and A. Edmonds and others},
    doi = {10.1103/PhysRevAccelBeams.20.030101},
    eprint = {1610.07850},
    journal = {Phys. Rev. Accel. Beams},
    pages = {030101},
    primaryclass = {physics.acc-ph},
    title = {Delivering the world's most intense muon beam},
    volume = {20},
    year = {2017}
}

@article{Chen:muoniumreview-2026,
    archiveprefix = {arXiv},
    author = {Siyuan Chen and Mingchen Sun and Jian Tang},
    eprint = {2601.15818},
    primaryclass = {hep-ex},
    title = {Muon beams towards muonium physics: progress and prospects},
    year = {2026}
}

@article{DalMaso:HIMB2023,
    author = {Dal Maso, G. and Gabard, A. and Haj Tahar, M. and others},
    doi = {10.1051/epjconf/202328201012},
    journal = {EPJ Web Conf.},
    pages = {01012},
    title = {Future facilities at {PSI}, the High-Intensity Muon Beams ({HIMB}) project},
    volume = {282},
    year = {2023}
}

@inproceedings{Pathak:PIPII2024,
    address = {Geneva, Switzerland},
    author = {Pathak, A. and Saini, A. and Pozdeyev, E.},
    booktitle = {Proceedings of the 15th International Particle Accelerator Conference ({IPAC}'24)},
    doi = {10.18429/JACoW-IPAC2024-MOPC30},
    pages = {125--128},
    publisher = {JACoW Publishing},
    title = {Final physics design of proton improvement {Plan-II} at {Fermilab}},
    year = {2024}
}

@misc{PSI:PiE5,
    author = {{Paul Scherrer Institute}},
    howpublished = {\url{https://www.psi.ch/en/sbl/pie5-beamline}},
    note = {Accessed: 29 October 2026},
    title = {{$\pi$E5} Beam Line}
}

@article{MEG:Polarization2016,
    author = {{MEG Collaboration}},
    doi = {10.1140/epjc/s10052-016-4047-3},
    journal = {Eur. Phys. J. C},
    pages = {223},
    title = {Muon polarization in the {MEG} experiment: predictions and measurements},
    volume = {76},
    year = {2016}
}

@article{MEGII:Design2018,
    archiveprefix = {arXiv},
    author = {A. M. Baldini and others},
    doi = {10.1140/epjc/s10052-018-5845-6},
    eprint = {1801.04688},
    journal = {Eur. Phys. J. C},
    pages = {380},
    primaryclass = {physics.ins-det},
    title = {The design of the {MEG II} experiment},
    volume = {78},
    year = {2018}
}

@article{Kreitzman:TRIUMF2018,
    author = {S. R. Kreitzman and G. D. Morris},
    doi = {10.7566/JPSCP.21.011056},
    journal = {JPS Conf. Proc.},
    pages = {011056},
    title = {{TRIUMF} {MuSR} and $\beta${NMR} Research Facilities},
    volume = {21},
    year = {2018}
}

@misc{Kreitzman:TRIUMF2025,
    author = {Sydney Kreitzman and others},
    howpublished = {Presentation at the 16th International Conference on Muon Spin Rotation, Relaxation and Resonance ({$\mu$SR2025}), 22 July 2025, \url{https://indico.triumf.ca/event/537/contributions/6793/}},
    title = {{TRIUMF CMMS} Facility Update},
    year = {2025}
}

@article{Cook:MuSICProduction2013,
    author = {S. Cook and R. D'Arcy and M. Fukuda and others},
    doi = {10.1088/1742-6596/408/1/012079},
    journal = {J. Phys.: Conf. Ser.},
    pages = {012079},
    title = {First measurements of muon production rate using a novel pion capture system at {MuSIC}},
    volume = {408},
    year = {2013}
}

@article{Abbon:COMPASS2007,
    author = {P. Abbon and others},
    doi = {10.1016/j.nima.2007.03.026},
    journal = {Nucl. Instrum. Methods Phys. Res. A},
    pages = {455--518},
    title = {The {COMPASS} experiment at {CERN}},
    volume = {577},
    year = {2007}
}

@article{Valetov:HIMBBeamline2024,
    author = {Valetov, Eremey and {Dal Maso}, Giovanni and Kettle, Peter-Raymond and Knecht, Andreas and Papa, Angela},
    doi = {10.3390/particles7030039},
    journal = {Particles},
    number = {3},
    pages = {683--691},
    title = {Beamline Optimisation for High-Intensity Muon Beams at {PSI} Using the Heterogeneous Island Model},
    volume = {7},
    year = {2024}
}

@article{Mu2e:RunISensitivity2023,
    archiveprefix = {arXiv},
    author = {F. Abdi and others},
    collaboration = {Mu2e},
    doi = {10.3390/universe9010054},
    eprint = {2210.11380},
    journal = {Universe},
    number = {1},
    pages = {54},
    primaryclass = {hep-ex},
    reportnumber = {FERMILAB-PUB-22-749-PPD},
    title = {{Mu2e} {Run I} sensitivity projections for the neutrinoless \(\mu^- \to e^-\) conversion search in aluminum},
    volume = {9},
    year = {2023}
}

@article{Chen:MELODYBeamline2023,
    author = {C. Chen and Y. Bao and N. Vassilopoulos},
    doi = {10.1088/1742-6596/2462/1/012027},
    journal = {J. Phys.: Conf. Ser.},
    pages = {012027},
    title = {Design of the surface muon beamline of {MELODY}},
    volume = {2462},
    year = {2023}
}

@article{Zhang:MELODYIntensity2025,
    author = {Mingzhi Zhang and Juan Liu and Ruirui Fan and others},
    doi = {10.1016/j.nima.2025.170871},
    journal = {Nucl. Instrum. Methods Phys. Res. A},
    pages = {170871},
    title = {Direct measurement of pulsed surface muon beam intensity
using plastic scintillator detector and current shape
discrimination method},
    volume = {1081},
    year = {2026}
}

@article{Sheng:HFRSOptics2024,
    author = {L. N. Sheng and X. H. Zhang and H. Ren and others},
    doi = {10.1016/j.nimb.2023.165214},
    journal = {Nucl. Instrum. Methods Phys. Res. B},
    pages = {165214},
    title = {Ion-optical updates and performance analysis of High energy FRagment Separator ({HFRS}) at {HIAF}},
    volume = {547},
    year = {2024}
}

@article{Cai:CiADS2024,
    archiveprefix = {arXiv},
    author = {Han-Jie Cai and Yuan He and Shuhui Liu and others},
    doi = {10.1103/PhysRevAccelBeams.27.023403},
    eprint = {2309.01520},
    journal = {Phys. Rev. Accel. Beams},
    pages = {023403},
    primaryclass = {physics.acc-ph},
    title = {Towards a high-intensity muon source},
    volume = {27},
    year = {2024}
}

@article{Liu:SHINE2025,
    archiveprefix = {arXiv},
    author = {Fangchao Liu and Yusuke Takeuchi and Si Chen and others},
    doi = {10.1103/t2d3-xqnp},
    eprint = {2503.01597},
    journal = {Phys. Rev. Accel. Beams},
    number = {8},
    pages = {083401},
    primaryclass = {physics.acc-ph},
    title = {Simulation studies of a high-repetition-rate electron-driven surface muon beamline at {SHINE}},
    volume = {28},
    year = {2025}
}

@article{Lv:SHINEMuonSource2023,
    archiveprefix = {arXiv},
    author = {M. Lv and J. Wang and K. Khaw},
    doi = {10.18429/JACoW-IPAC2023-TUPA087},
    eprint = {2307.01455},
    journal = {Proc. IPAC'23},
    pages = {1522--1525},
    primaryclass = {physics.acc-ph},
    title = {A pulsed muon source based on a high-repetition-rate electron accelerator},
    year = {2023}
}

@article{Bao:MELODYProgress2023,
    author = {Y. Bao and C. Chen and N. Vassilopoulos and others},
    doi = {10.1088/1742-6596/2462/1/012034},
    journal = {J. Phys.: Conf. Ser.},
    pages = {012034},
    title = {Progress report on Muon Source Project at {CSNS}},
    volume = {2462},
    year = {2023}
}

@article{An:HuizhouPrecision2025,
    archiveprefix = {arXiv},
    author = {Fengpeng An and Dong Bai and Hanjie Cai and Siyuan Chen and Xurong Chen and Hongyue Duyang and Leyun Gao and others},
    doi = {10.1088/0256-307X/42/11/110102},
    eprint = {2504.21050},
    journal = {Chinese Physics Letters},
    number = {11},
    pages = {110102},
    primaryclass = {hep-ex},
    title = {High-Precision Physics Experiments at {Huizhou} Large-Scale Scientific Facilities},
    volume = {42},
    year = {2025}
}

@misc{He:NuFact2026,
    author = {He, Zhengyang},
    howpublished = {Presentation at {NuFact 2026}, Tsung-Dao Lee Institute, Shanghai, 1 September 2026, \url{https://indico-tdli.sjtu.edu.cn/event/4693/contributions/20214/}},
    title = {Status and future plans for the {Muon Source Program} at {HIAF}},
    year = {2026}
}

@article{Zhang:MuonDMBeam2026,
    archiveprefix = {arXiv},
    author = {Rongfeng Zhang and {C.-E. Liu} and Ruihu Zhu and others},
    eprint = {2609.11617},
    primaryclass = {hep-ex},
    title = {Projected Sensitivity to Slow Muonphilic Dark Matter with Accelerator Muon Beams},
    year = {2026}
}

@article{Achenbach:JLabBeamDump2025,
    archiveprefix = {arXiv},
    author = {Patrick Achenbach and Andrei Afanasev and Pawel Ambrozewicz and others},
    doi = {10.1140/epja/s10050-025-01748-6},
    eprint = {2510.09652},
    journal = {Eur. Phys. J. A},
    pages = {285},
    primaryclass = {physics.acc-ph},
    title = {A Beamdump Facility at {Jefferson Lab}},
    volume = {61},
    year = {2025}
}

@article{Williams:SEEMS2022,
    archiveprefix = {arXiv},
    author = {Travis J. Williams and Gregory J. MacDougall and Bernie W. Riemer and others},
    doi = {10.1063/5.0135721},
    eprint = {2212.09823},
    journal = {Rev. Sci. Instrum.},
    number = {3},
    pages = {033908},
    primaryclass = {physics.acc-ph},
    title = {{SEEMS}: A Single Event Effects and Muon Spectroscopy facility at the {Spallation Neutron Source}},
    volume = {94},
    year = {2023}
}

@article{Choi:RAONMuon2014,
    archiveprefix = {arXiv},
    author = {Suyong Choi and Jeongwon Park and Youn Jung Roh},
    doi = {10.3938/jkps.66.762},
    eprint = {1406.2091},
    journal = {J. Korean Phys. Soc.},
    number = {5},
    pages = {762--767},
    primaryclass = {physics.acc-ph},
    title = {The design of an optimized muon beamline},
    volume = {66},
    year = {2015}
}

@article{Palo:CLFV2025,
    archiveprefix = {arXiv},
    author = {Dylan Palo},
    eprint = {2505.04764},
    primaryclass = {hep-ex},
    title = {Charged Lepton Flavor Violating Experiments with Muons},
    year = {2025}
}

@misc{Fermilab:MuonG2Photo2021,
    author = {{Fermi National Accelerator Laboratory}},
    howpublished = {Fermilab Newsroom, \url{https://news.fnal.gov/2021/04/first-results-from-fermilabs-muon-g-2-experiment-strengthen-evidence-of-new-physics/}},
    month = {April},
    note = {Photograph by Reidar Hahn; published 7 April 2021; accessed 22 July 2026},
    title = {First results from {Fermilab}'s muon $g-2$ experiment strengthen evidence of new physics},
    year = {2021}
}

@article{Abe:JPARCG2EDM2019,
    archiveprefix = {arXiv},
    author = {M. Abe and S. Bae and G. Beer and others},
    doi = {10.1093/ptep/ptz030},
    eprint = {1901.03047},
    journal = {Prog. Theor. Exp. Phys.},
    pages = {053C02},
    primaryclass = {physics.ins-det},
    title = {A new approach for measuring the muon anomalous magnetic moment and electric dipole moment},
    volume = {2019},
    year = {2019}
}

@article{Han:DoublyChargedHiggs2021,
    archiveprefix = {arXiv},
    author = {Chengcheng Han and Da Huang and Jian Tang and Yu Zhang},
    doi = {10.1103/PhysRevD.103.055023},
    eprint = {2102.00758},
    journal = {Physical Review D},
    pages = {055023},
    primaryclass = {hep-ph},
    title = {Probing the Doubly Charged {Higgs} Boson with a Muonium to Antimuonium Conversion Experiment},
    volume = {103},
    year = {2021}
}

@article{Zhao:Aerogel2024,
    archiveprefix = {arXiv},
    author = {Shihan Zhao and Jian Tang},
    doi = {10.1103/PhysRevD.109.072012},
    eprint = {2401.00222},
    journal = {Physical Review D},
    number = {7},
    pages = {072012},
    primaryclass = {hep-ex},
    title = {Optimization of muonium yield in perforated silica aerogel},
    volume = {109},
    year = {2024}
}

@article{Lu:MACEPTS2025,
    archiveprefix = {arXiv},
    author = {Guihao Lu and Shihan Zhao and Siyuan Chen and Jian Tang},
    doi = {10.1103/2yb7-zv76},
    eprint = {2508.07922},
    journal = {Physical Review Accelerators and Beams},
    pages = {031602},
    primaryclass = {hep-ex},
    title = {Positron Transport System for Muonium-to-Antimuonium Conversion Experiment},
    volume = {29},
    year = {2026}
}

@article{Chen:MACECalorimeter2024,
    archiveprefix = {arXiv},
    author = {Siyuan Chen and Shihan Zhao and Weizhi Xiong and Ye Tian and Hui Jiang and Jiacheng Ling and Shishe Wang and Jian Tang},
    doi = {10.15302/frontphys.2025.035202},
    eprint = {2408.17114},
    journal = {Frontiers of Physics},
    number = {3},
    pages = {035202},
    primaryclass = {physics.ins-det},
    title = {Design of a {CsI(Tl)} Calorimeter for Muonium-to-Antimuonium Conversion Experiment},
    volume = {20},
    year = {2025}
}

@article{Willmann:MACS1999,
    archiveprefix = {arXiv},
    author = {L. Willmann and P. V. Schmidt and H. P. Wirtz and others},
    doi = {10.1103/PhysRevLett.82.49},
    eprint = {hep-ex/9807011},
    journal = {Phys. Rev. Lett.},
    pages = {49--52},
    title = {New Bounds from Searching for Muonium to Antimuonium Conversion},
    volume = {82},
    year = {1999}
}

@article{Zhao:MACEProgress2024,
    author = {Shihan Zhao and Jian Tang},
    doi = {10.1016/j.nuclphysbps.2024.05.005},
    journal = {Nuclear and Particle Physics Proceedings},
    pages = {24--28},
    title = {Progress of muonium-to-antimuonium conversion experiment},
    volume = {345},
    year = {2024}
}

@article{Bertl:SINDRUMII2006,
    author = {W. Bertl and others},
    doi = {10.1140/epjc/s2006-02582-x},
    journal = {Eur. Phys. J. C},
    number = {2},
    pages = {337--346},
    title = {A search for {$\mu$-$e$} conversion in muonic gold},
    volume = {47},
    year = {2006}
}

@article{Ricci:Mu2e2025,
    archiveprefix = {arXiv},
    author = {{A. M. Ricci (for the Mu2e Collaboration)}},
    doi = {10.1051/epjconf/202533701057},
    eprint = {2501.05492},
    journal = {EPJ Web Conf.},
    pages = {01057},
    primaryclass = {hep-ex},
    title = {Multi-track reconstruction algorithms in the {Mu2e} experiment},
    volume = {337},
    year = {2025}
}

@article{Fujii:COMET2023,
    archiveprefix = {arXiv},
    author = {{Y. Fujii (for the COMET Collaboration)}},
    doi = {10.1088/1748-0221/18/10/C10010},
    eprint = {2308.14275},
    journal = {JINST},
    pages = {C10010},
    primaryclass = {hep-ex},
    title = {A search for a muon to electron conversion in {COMET}},
    volume = {18},
    year = {2023}
}

@article{Oishi:COMETRangeCounter2025,
    archiveprefix = {arXiv},
    author = {Kou Oishi and Masaharu Aoki and Shion Kuribayashi and others},
    doi = {10.1016/j.nima.2025.170904},
    eprint = {2505.07464},
    journal = {Nucl. Instrum. Methods Phys. Res. A},
    pages = {170904},
    primaryclass = {physics.ins-det},
    title = {Development of the Range Counter for the {COMET} Phase-$\alpha$ Experiment},
    volume = {1082},
    year = {2026}
}

@article{Xu:COMETMBM2024,
    author = {Yu Xu and Yun-Song Ning and Zhi-Zhen Qin and Yao Teng and Chang-Qing Feng and Jian Tang and Yu Chen and Yoshinori Fukao and Satoshi Mihara and Kou Oishi},
    doi = {10.1007/s41365-024-01442-0},
    journal = {Nuclear Science and Techniques},
    pages = {79},
    title = {Development of a Scintillating-Fiber-Based Beam Monitor for the Coherent Muon-to-Electron Transition Experiment},
    volume = {35},
    year = {2024}
}

@article{Moritsu:COMET2022,
    author = {M. Moritsu},
    doi = {10.3390/universe8040196},
    journal = {Universe},
    pages = {196},
    title = {Search for Muon-to-Electron Conversion with the {COMET} Experiment},
    volume = {8},
    year = {2022}
}

@article{Bellgardt:SINDRUM1988,
    author = {U. Bellgardt and G. Otter and R. Eichler and others},
    doi = {10.1016/0550-3213(88)90462-2},
    journal = {Nucl. Phys. B},
    pages = {1--6},
    title = {Search for the Decay $\mu^+ \to e^+ e^+ e^-$},
    volume = {299},
    year = {1988}
}

@article{Amarinei:Mu3e2025,
    archiveprefix = {arXiv},
    author = {Robert Mihai Amarinei},
    eprint = {2501.14667},
    primaryclass = {hep-ex},
    title = {The {Mu3e} Experiment: Status and Short-Term Plans},
    year = {2025}
}

@article{Aiba:HIMB2021,
    archiveprefix = {arXiv},
    author = {M. Aiba and others},
    eprint = {2111.05788},
    primaryclass = {hep-ex},
    title = {Science Case for the new High-Intensity Muon Beams {HIMB} at {PSI}},
    year = {2021}
}

@article{Cattaneo:FutureMEGLoI2025,
    archiveprefix = {arXiv},
    author = {P. W. Cattaneo and W. Ootani and F. Renga and A. Sch\"oning and others},
    eprint = {2504.18831},
    primaryclass = {hep-ex},
    title = {Letter of Intent for a future $\mu^+\to e^+\gamma$ experiment at the {High Intensity Muon Beam} facility at {PSI}},
    year = {2025}
}

@misc{Chen:HistoryPlot,
    author = {S. Chen},
    howpublished = {\url{https://github.com/phd-csy/muon-clfv-history-plot}},
    title = {Muon {CLFV} History Plot},
    year = {2026}
}

@techreport{Grange:MuonG2TDR2015,
    address = {Batavia, Illinois},
    archiveprefix = {arXiv},
    author = {J. Grange and others},
    doi = {10.2172/1251172},
    eprint = {1501.06858},
    institution = {Fermi National Accelerator Laboratory},
    month = {January},
    number = {FERMILAB-FN-0992-E, FERMILAB-DESIGN-2014-02},
    primaryclass = {physics.ins-det},
    title = {Muon $g-2$ Technical Design Report},
    year = {2015}
}

@article{MuonG2:Measurement2025,
    archiveprefix = {arXiv},
    author = {{Muon $g-2$ Collaboration}},
    doi = {10.1103/7clf-sm2v},
    eprint = {2506.03069},
    journal = {Phys. Rev. Lett.},
    number = {10},
    pages = {101802},
    primaryclass = {hep-ex},
    title = {Measurement of the Positive Muon Anomalous Magnetic Moment to 127 ppb},
    volume = {135},
    year = {2025}
}

@article{MuonG2:Final2026,
    archiveprefix = {arXiv},
    author = {Aguillard, D. P. and others},
    collaboration = {Muon $g-2$},
    eprint = {2606.17323},
    primaryclass = {hep-ex},
    title = {Final report on the measurement of the positive muon anomalous magnetic moment at {Fermilab} to 127 ppb},
    year = {2026}
}

@article{MuonG2:EDM2026,
    archiveprefix = {arXiv},
    author = {Aguillard, D. P. and others},
    collaboration = {Muon $g-2$},
    doi = {10.48550/arXiv.2608.11124},
    eprint = {2608.11124},
    primaryclass = {hep-ex},
    title = {An improved direct limit on the muon electric dipole moment},
    year = {2026}
}

@article{JPARC:g2EDMStatus2025,
    archiveprefix = {arXiv},
    author = {Graziano Venanzoni},
    doi = {10.22323/1.485.0427},
    eprint = {2512.20335},
    journal = {PoS},
    pages = {427},
    primaryclass = {hep-ex},
    title = {Status of the Muon $g-2$/{EDM} Experiment at {J-PARC}},
    volume = {EPS-HEP2025},
    year = {2026}
}

@article{Adelmann:muEDM2025,
    archiveprefix = {arXiv},
    author = {A. Adelmann and A. R. Bainbridge and I. Bailey and others},
    doi = {10.1140/epjc/s10052-025-14295-7},
    eprint = {2501.18979},
    journal = {Eur. Phys. J. C},
    pages = {622},
    primaryclass = {hep-ex},
    title = {A compact frozen-spin trap for the search for the electric dipole moment of the muon},
    volume = {85},
    year = {2025}
}

@article{Kanda:MuSEUM2020,
    archiveprefix = {arXiv},
    author = {S. Kanda and Y. Fukao and Y. Ikedo and others},
    doi = {10.1016/j.physletb.2021.136154},
    eprint = {2004.05862},
    journal = {Phys. Lett. B},
    pages = {136154},
    primaryclass = {hep-ex},
    title = {New precise spectroscopy of the hyperfine structure in muonium with a high-intensity pulsed muon beam},
    volume = {815},
    year = {2021}
}

@article{Nishimura:MuSEUM2021,
    archiveprefix = {arXiv},
    author = {S. Nishimura and others},
    doi = {10.1103/PhysRevA.104.L020801},
    eprint = {2007.12386},
    journal = {Phys. Rev. A},
    pages = {L020801},
    primaryclass = {hep-ex},
    title = {Rabi-oscillation spectroscopy of the hyperfine structure of muonium atoms},
    volume = {104},
    year = {2021}
}

@article{Cortinovis:MuMASS2023,
    archiveprefix = {arXiv},
    author = {Irene Cortinovis and Ben Ohayon and Lucas de Sousa Borges and others},
    doi = {10.1140/epjd/s10053-023-00639-z},
    eprint = {2301.12883},
    journal = {Eur. Phys. J. D},
    pages = {66},
    primaryclass = {physics.atom-ph},
    title = {Update of muonium  {$1S$--$2S$} transition frequency},
    volume = {77},
    year = {2023}
}

@article{Janka:Muonium2S2P2022,
    archiveprefix = {arXiv},
    author = {G. Janka and B. Ohayon and I. Cortinovis and others},
    doi = {10.1038/s41467-022-34672-0},
    eprint = {2205.06202},
    journal = {Nature Commun.},
    pages = {7273},
    primaryclass = {physics.atom-ph},
    title = {Measurement of the transition frequency from
{$2S_{1/2}$}, {$F=0$} to {$2P_{1/2}$}, {$F=1$} states in muonium},
    volume = {13},
    year = {2022}
}

@article{Pohl:ProtonRadius2010,
    author = {R. Pohl and A. Antognini and F. Nez and others},
    doi = {10.1038/nature09250},
    journal = {Nature},
    pages = {213--216},
    title = {The size of the proton},
    volume = {466},
    year = {2010}
}

@article{Ohayon:MuonicXRays2023,
    archiveprefix = {arXiv},
    author = {Ben Ohayon and Andreas Abeln and Silvia Bara and others},
    doi = {10.3390/physics6010015},
    eprint = {2310.03846},
    journal = {Physics},
    number = {1},
    pages = {206--215},
    primaryclass = {nucl-ex},
    title = {Towards Precision Muonic X-Ray Measurements of Charge Radii of Light Nuclei},
    volume = {6},
    year = {2024}
}

@article{Xiong:ProtonRadiusReview2023,
    archiveprefix = {arXiv},
    author = {Weizhi Xiong and Chao Peng},
    doi = {10.3390/universe9040182},
    eprint = {2302.13818},
    journal = {Universe},
    number = {4},
    pages = {182},
    primaryclass = {nucl-ex},
    title = {Proton Electric Charge Radius from Lepton Scattering},
    volume = {9},
    year = {2023}
}

@article{Amaro:HyperMu2022,
    archiveprefix = {arXiv},
    author = {P. Amaro and A. Adamczak and M. {Abdou Ahmed} and others},
    doi = {10.21468/SciPostPhys.13.2.020},
    eprint = {2112.00138},
    journal = {SciPost Phys.},
    pages = {020},
    primaryclass = {physics.atom-ph},
    title = {Laser excitation of the 1s-hyperfine transition in muonic hydrogen},
    volume = {13},
    year = {2022}
}

@article{Bonesini:FAMUDet2025,
    archiveprefix = {arXiv},
    author = {M. Bonesini},
    doi = {10.1016/j.nima.2025.170780},
    eprint = {2507.01819},
    journal = {Nucl. Instrum. Methods Phys. Res. A},
    pages = {170780},
    primaryclass = {physics.ins-det},
    title = {The fast {X-ray} detector system of the {FAMU} experiment at {RAL}},
    volume = {1080},
    year = {2025}
}

@article{FAMU:Operation2025,
    archiveprefix = {arXiv},
    author = {{FAMU Collaboration}},
    doi = {10.1140/epja/s10050-025-01758-4},
    eprint = {2509.10350},
    journal = {Eur. Phys. J. A},
    pages = {284},
    primaryclass = {physics.ins-det},
    title = {First operation of the {FAMU} experiment at the {RIKEN-RAL} high intensity muon beam facility},
    volume = {61},
    year = {2025}
}

@article{Okumura:MuonicNeon2023,
    author = {Okumura, T. and Azuma, T. and Bennett, D. A. and Chiu, I. and Doriese, W. B. and Durkin, M. S. and Fowler, J. W. and Gard, J. D. and Hashimoto, T. and Hayakawa, R. and Hilton, G. C. and Ichinohe, Y. and Indelicato, P. and Isobe, T. and Kanda, S. and Katsuragawa, M. and Kawamura, N. and Kino, Y. and Mine, K. and Miyake, Y. and Morgan, K. M. and Ninomiya, K. and Noda, H. and O'Neil, G. C. and Okada, S. and Okutsu, K. and Paul, N. and Reintsema, C. D. and Schmidt, D. R. and Shimomura, K. and Strasser, P. and Suda, H. and Swetz, D. S. and Takahashi, T. and Takeda, S. and Takeshita, S. and Tampo, M. and Tatsuno, H. and Ueno, Y. and Ullom, J. N. and Watanabe, S. and Yamada, S.},
    doi = {10.1103/PhysRevLett.130.173001},
    issue = {17},
    journal = {Phys. Rev. Lett.},
    month = {Apr},
    numpages = {7},
    pages = {173001},
    publisher = {American Physical Society},
    title = {Proof-of-Principle Experiment for Testing Strong-Field Quantum Electrodynamics with Exotic Atoms: High Precision X-Ray Spectroscopy of Muonic Neon},
    url = {https://link.aps.org/doi/10.1103/PhysRevLett.130.173001},
    volume = {130},
    year = {2023}
}

@article{Eizenberg:MuonicBe2026,
    archiveprefix = {arXiv},
    author = {Eizenberg, Ofir and others},
    eprint = {2607.13690},
    primaryclass = {nucl-ex},
    title = {Nuclear Charge Radius of {$^9$Be} from Muonic Atom Spectroscopy Using a Microcalorimeter},
    year = {2026}
}

@article{Adamczak:muX2023,
    archiveprefix = {arXiv},
    author = {A. Adamczak and A. Antognini and N. Berger and others},
    doi = {10.1140/epja/s10050-023-00930-y},
    eprint = {2209.14365},
    journal = {Eur. Phys. J. A},
    pages = {15},
    primaryclass = {physics.ins-det},
    title = {Muonic atom spectroscopy with microgram target material},
    volume = {59},
    year = {2023}
}

@article{Charpak:MuonG21965,
    author = {G. Charpak and F. J. M. Farley and R. L. Garwin and others},
    doi = {10.1007/BF02783344},
    journal = {Nuovo Cim.},
    pages = {1241--1363},
    title = {The anomalous magnetic moment of the muon},
    volume = {37},
    year = {1965}
}

@article{Bailey:MuonG21968,
    author = {J. Bailey and W. Bartl and G. von Bochmann and others},
    doi = {10.1016/0370-2693(68)90261-X},
    journal = {Phys. Lett. B},
    pages = {287--290},
    title = {Precision measurement of the anomalous magnetic moment of the muon},
    volume = {28},
    year = {1968}
}

@article{Bailey:MuonG21979,
    author = {J. Bailey and K. Borer and F. Combley and others},
    doi = {10.1016/0550-3213(79)90292-X},
    journal = {Nucl. Phys. B},
    pages = {1--75},
    title = {Final report on the {CERN} muon storage ring including the anomalous magnetic moment and the electric dipole moment of the muon, and a direct test of relativistic time dilation},
    volume = {150},
    year = {1979}
}

@article{Bennett:E8212006,
    archiveprefix = {arXiv},
    author = {G. W. Bennett and others},
    doi = {10.1103/PhysRevD.73.072003},
    eprint = {hep-ex/0602035},
    journal = {Phys. Rev. D},
    pages = {072003},
    title = {Final Report of the {E821} Muon Anomalous Magnetic Moment Measurement at {BNL}},
    volume = {73},
    year = {2006}
}

@article{Abi:MuonG22021,
    archiveprefix = {arXiv},
    author = {B. Abi and others},
    doi = {10.1103/PhysRevLett.126.141801},
    eprint = {2104.03281},
    journal = {Phys. Rev. Lett.},
    pages = {141801},
    title = {Measurement of the Positive Muon Anomalous Magnetic Moment to 0.46 ppm},
    volume = {126},
    year = {2021}
}

@article{Aguillard:MuonG22023,
    archiveprefix = {arXiv},
    author = {D. P. Aguillard and others},
    doi = {10.1103/PhysRevLett.131.161802},
    eprint = {2308.06230},
    journal = {Phys. Rev. Lett.},
    pages = {161802},
    title = {Measurement of the Positive Muon Anomalous Magnetic Moment to 0.20 ppm},
    volume = {131},
    year = {2023}
}

@article{Berley:MuonEDM1958,
    author = {D. Berley and R. L. Garwin and G. Gidal and L. M. Lederman},
    doi = {10.1103/PhysRevLett.1.144},
    journal = {Phys. Rev. Lett.},
    pages = {144--146},
    title = {Electric Dipole Moment of the Muon},
    volume = {1},
    year = {1958}
}

@article{Bailey:MuonEDM1978,
    author = {J. Bailey and K. Borer and F. Combley and others},
    doi = {10.1088/0305-4616/4/3/010},
    journal = {J. Phys. G},
    pages = {345--352},
    title = {New limits on the electric dipole moment of positive and negative muons},
    volume = {4},
    year = {1978}
}

@article{Bennett:MuonEDM2009,
    archiveprefix = {arXiv},
    author = {G. W. Bennett and others},
    doi = {10.1103/PhysRevD.80.052008},
    eprint = {0811.1207},
    journal = {Phys. Rev. D},
    pages = {052008},
    title = {An improved limit on the muon electric dipole moment},
    volume = {80},
    year = {2009}
}

@article{Favart:MuoniumHFS1973,
    author = {D. Favart and P. M. McIntyre and D. Y. Stowell and others},
    doi = {10.1103/PhysRevA.8.1195},
    journal = {Phys. Rev. A},
    pages = {1195--1218},
    title = {Precision Experiments on Muonium. {III}. {Ramsey} Resonance in Zero Field},
    volume = {8},
    year = {1973}
}

@article{Casperson:MuoniumHFS1975,
    author = {D. E. Casperson and T. W. Crane and V. W. Hughes and others},
    doi = {10.1016/0370-2693(75)90099-4},
    journal = {Phys. Lett. B},
    pages = {397--400},
    title = {A new high precision measurement of the muonium hyperfine structure interval {$\Delta\nu$}},
    volume = {59},
    year = {1975}
}

@article{Mariam:MuoniumHFS1982,
    author = {F. G. Mariam and W. Beer and P. R. Bolton and others},
    doi = {10.1103/PhysRevLett.49.993},
    journal = {Phys. Rev. Lett.},
    pages = {993--996},
    title = {Higher Precision Measurement of the {HFS} Interval of Muonium and of the Muon Magnetic Moment},
    volume = {49},
    year = {1982}
}

@article{Liu:MuoniumHFS1999,
    author = {W. Liu and M. G. Boshier and S. Dhawan and others},
    doi = {10.1103/PhysRevLett.82.711},
    journal = {Phys. Rev. Lett.},
    pages = {711--714},
    title = {High Precision Measurements of the Ground State Hyperfine Structure Interval of Muonium and of the Muon Magnetic Moment},
    volume = {82},
    year = {1999}
}

@article{Jungmann:Muonium1S2S1991,
    author = {K. Jungmann and P. E. G. Baird and J. R. M. Barr and others},
    doi = {10.1007/BF01426380},
    journal = {Z. Phys. D},
    pages = {241--243},
    title = {Two-photon laser spectroscopy of the muonium {1S--2S} transition},
    volume = {21},
    year = {1991}
}

@article{Maas:Muonium1S2S1994,
    author = {F. E. Maas and B. Braun and H. Geerds and others},
    doi = {10.1016/0375-9601(94)90903-2},
    journal = {Phys. Lett. A},
    pages = {247--254},
    title = {A measurement of the {1S--2S} transition frequency in muonium},
    volume = {187},
    year = {1994}
}

@article{Meyer:Muonium1S2S2000,
    archiveprefix = {arXiv},
    author = {V. Meyer and S. N. Bagayev and P. E. G. Baird and others},
    doi = {10.1103/PhysRevLett.84.1136},
    eprint = {hep-ex/9907013},
    journal = {Phys. Rev. Lett.},
    pages = {1136--1139},
    title = {Measurement of the {1S--2S} Energy Interval in Muonium},
    volume = {84},
    year = {2000}
}

@article{Cline:PiM1MUSE2022,
    archiveprefix = {arXiv},
    author = {E. Cline and W. Lin and P. Roy and others},
    doi = {10.1103/PhysRevC.105.055201},
    eprint = {2109.09508},
    journal = {Phys. Rev. C},
    pages = {055201},
    title = {Characterization of muon and electron beams in the {Paul Scherrer Institute} {PiM1} channel for the {MUSE} experiment},
    volume = {105},
    year = {2022}
}

@article{Li:MUSERadiative2024,
    archiveprefix = {arXiv},
    author = {L. Li and S. Strauch and J. C. Bernauer and others},
    doi = {10.1140/epja/s10050-023-01215-0},
    eprint = {2307.06417},
    journal = {Eur. Phys. J. A},
    pages = {8},
    title = {Instrumental uncertainties in radiative corrections for the {MUSE} experiment},
    volume = {60},
    year = {2024}
}

@article{Lin:MUSECalorimeter2025,
    archiveprefix = {arXiv},
    author = {W. Lin and T. Rostomyan and R. Gilman and others},
    doi = {10.1016/j.nima.2025.170754},
    eprint = {2408.13380},
    journal = {Nucl. Instrum. Methods Phys. Res. A},
    pages = {170754},
    title = {The {MUSE} beamline calorimeter},
    volume = {1080},
    year = {2025}
}

@article{Gurgone:MUonETheory2024,
    archiveprefix = {arXiv},
    author = {{A. Gurgone (on behalf of the MUonE Collaboration)}},
    doi = {10.22323/1.449.0307},
    eprint = {2401.06491},
    journal = {PoS},
    pages = {307},
    primaryclass = {hep-ph},
    title = {Theory for the {MUonE} experiment},
    volume = {EPS-HEP2023},
    year = {2024}
}

@article{Ruzi:PKMu2023,
    archiveprefix = {arXiv},
    author = {Alim Ruzi and Chen Zhou and Xiaohu Sun and others},
    doi = {10.1142/S0217751X23501543},
    eprint = {2303.18117},
    journal = {International Journal of Modern Physics A},
    number = {29n30},
    pages = {2350154},
    primaryclass = {hep-ex},
    title = {Probing dark matter using free leptons: {PKMUON}},
    volume = {38},
    year = {2023}
}

@article{Gao:PKMuOverview2025,
    archiveprefix = {arXiv},
    author = {Leyun Gao and {C.-E. Liu} and Qite Li and others},
    doi = {10.1142/S0217732325300083},
    eprint = {2503.22956},
    journal = {Modern Physics Letters A},
    number = {24},
    pages = {2530008},
    primaryclass = {hep-ex},
    title = {Probing and knocking with muons},
    volume = {40},
    year = {2025}
}

@article{GilDominguez:MUSERadius2023,
    archiveprefix = {arXiv},
    author = {F. Gil-Dom{\'i}nguez and J. M. Alarc{\'o}n and C. Weiss},
    doi = {10.1103/PhysRevD.108.074026},
    eprint = {2306.01037},
    journal = {Phys. Rev. D},
    pages = {074026},
    primaryclass = {hep-ph},
    title = {Proton charge radius extraction from muon scattering at {MUSE} using dispersively improved chiral effective field theory},
    volume = {108},
    year = {2023}
}

@article{Rostomyan:MUSEtiming2021,
    archiveprefix = {arXiv},
    author = {T. Rostomyan and E. Cline and I. Lavrukhin and others},
    doi = {10.1016/j.nima.2020.164801},
    eprint = {2007.12207},
    journal = {Nucl. Instrum. Methods Phys. Res. A},
    pages = {164801},
    title = {Timing detectors with {SiPM} read-out for the {MUSE} experiment at {PSI}},
    volume = {986},
    year = {2021}
}

@article{Roy:MUSEHydrogen2020,
    archiveprefix = {arXiv},
    author = {P. Roy and S. Corsetti and M. Dimond and others},
    doi = {10.1016/j.nima.2019.162874},
    eprint = {1907.03022},
    journal = {Nucl. Instrum. Methods Phys. Res. A},
    pages = {162874},
    title = {A liquid hydrogen target for the {MUSE} experiment at {PSI}},
    volume = {949},
    year = {2020}
}

@article{Ratvasky:MUSEPostVeto2026,
    archiveprefix = {arXiv},
    author = {R. Ratvasky and T. Rostomyan and M. Ali and others},
    doi = {10.1016/j.nima.2026.171621},
    eprint = {2602.10279},
    journal = {Nucl. Instrum. Methods Phys. Res. A},
    pages = {171621},
    primaryclass = {physics.ins-det},
    title = {The {MUSE} target chamber post veto},
    volume = {1090},
    year = {2026}
}

@techreport{MUonE:PhaseI2024,
    address = {Geneva},
    author = {{MUonE Collaboration}},
    institution = {CERN},
    number = {CERN-SPSC-2024-015, SPSC-P-370},
    title = {Proposal for {Phase I} of the {MUonE} experiment},
    year = {2024}
}

@article{Pesaresi:MUonEPhaseI2026,
    author = {{M. Pesaresi, on behalf of the MUonE Collaboration}},
    doi = {10.1088/1748-0221/21/04/C04022},
    journal = {JINST},
    pages = {C04022},
    title = {Demonstration of a small-scale version of the {MUonE} experiment},
    volume = {21},
    year = {2026}
}

@article{Abbiendi:MUonETestBeam2021,
    archiveprefix = {arXiv},
    author = {G. Abbiendi and others},
    doi = {10.1088/1748-0221/16/06/P06005},
    eprint = {2102.11111},
    journal = {JINST},
    pages = {P06005},
    primaryclass = {hep-ex},
    title = {A study of muon-electron elastic scattering in a test beam},
    volume = {16},
    year = {2021}
}

@article{Holst:MuonG2ThermalDM2022,
    archiveprefix = {arXiv},
    author = {Ian Holst and Dan Hooper and Gordan Krnjaic},
    doi = {10.1103/PhysRevLett.128.141802},
    eprint = {2107.09067},
    journal = {Phys. Rev. Lett.},
    pages = {141802},
    primaryclass = {hep-ph},
    title = {Simplest and Most Predictive Model of Muon $g-2$ and Thermal Dark Matter},
    volume = {128},
    year = {2022}
}

@techreport{NA64:Status2025,
    address = {Geneva},
    author = {P. Crivelli and L. Molina-Bueno and S. Gninenko and A. Celentano and L. Marsicano and V. Poliakov},
    collaboration = {NA64},
    institution = {CERN},
    number = {CERN-SPSC-2025-031, SPSC-SR-367},
    title = {{NA64} Status Report 2025},
    year = {2025}
}

@article{Liu:PKMuCosmicDM2026,
    archiveprefix = {arXiv},
    author = {{C.-E. Liu} and Rongfeng Zhang and Zijian Wang and others},
    doi = {10.1103/5jh7-fxf4},
    eprint = {2507.23458},
    journal = {Phys. Rev. Lett.},
    pages = {151001},
    primaryclass = {hep-ex},
    title = {Probing Cosmic Ray Composition and Muonphilic Dark Matter via Muon Tomography},
    volume = {136},
    year = {2026}
}

@article{Yu:PKMuTomography2024,
    archiveprefix = {arXiv},
    author = {Xudong Yu and Zijian Wang and {C.-E. Liu} and others},
    doi = {10.1103/PhysRevD.110.016017},
    eprint = {2402.13483},
    journal = {Phys. Rev. D},
    pages = {016017},
    primaryclass = {hep-ex},
    title = {Proposed {Peking University} muon experiment for muon tomography and dark matter search},
    volume = {110},
    year = {2024}
}

@article{Sauli:GEM1997,
    author = {F. Sauli},
    doi = {10.1016/S0168-9002(96)01172-2},
    journal = {Nucl. Instrum. Methods Phys. Res. A},
    pages = {531--534},
    title = {{GEM}: A new concept for electron amplification in gas detectors},
    volume = {386},
    year = {1997}
}

@article{Riegler:RPC2004,
    author = {W. Riegler and C. Lippmann},
    doi = {10.1016/j.nima.2003.10.031},
    journal = {Nucl. Instrum. Methods Phys. Res. A},
    pages = {86--90},
    title = {The physics of resistive plate chambers},
    volume = {518},
    year = {2004}
}

@article{Yang:DetectorPositionCalibration2026,
    author = {Zhe Yang and {C.-E. Liu} and Rongfeng Zhang and others},
    doi = {10.20173/j.cnki.ned.20260320.001},
    journal = {Nuclear Electronics \& Detection Technology},
    number = {4},
    pages = {561--568},
    title = {Research on Detector Position Correction Algorithm for Muon Scattering Detection System},
    volume = {46},
    year = {2026}
}

@article{Qin:ProjectionShiftedMuonTomography2025,
    archiveprefix = {arXiv},
    author = {Zibo Qin and Qite Li and Rongfeng Zhang and Pei Yu and {C.-E. Liu} and Liangwen Chen and Feng Zhang and Qiang Li},
    eprint = {2512.19747},
    primaryclass = {physics.ins-det},
    title = {Projection-shifted particle-flow imaging with cosmic-ray muons},
    year = {2025}
}

@article{Zhao:RawHitMuonTomography2026,
    archiveprefix = {arXiv},
    author = {Zhizheng Zhao and Changhao Qin and Rongfeng Zhang and Zibo Qin and Ziyu Xu and Xingyu Xiao and Qiang Li and Qite Li},
    eprint = {2606.20180},
    primaryclass = {physics.ins-det},
    title = {Nuisance-Aware Muon Tomography},
    year = {2026}
}

@article{MUonE:ECAL2024,
    author = {{E. Spedicato (for the MUonE Collaboration)}},
    doi = {10.1088/1748-0221/19/02/C02044},
    journal = {JINST},
    number = {02},
    pages = {C02044},
    title = {A prototype electromagnetic calorimeter for the {MUonE} experiment: status and first performance results},
    volume = {19},
    year = {2024}
}
\end{document}